\documentclass[floatfix,preprint,eqsecnum,nofootinbib,amsmath,amssymb]{revtex4}

\newcommand\mystartdate{14 September 2011}

\newcommand\mydates{\mystartdate}

\usepackage{graphicx} 
\usepackage{bm}       

\newcommand{\bodyskip}{\baselineskip 18pt plus 1pt minus 1pt}

\newcommand{\footnoteskip}{\baselineskip 12pt plus 1pt minus 1pt}
\newcommand{\abstractskip}{\baselineskip 13pt plus 1pt minus 1pt}

\newcommand{\captionskip}{\footnotesize \baselineskip 12pt plus 1pt minus 1pt}

\newcommand{\smI}{{\rm\scriptscriptstyle I}}
\newcommand{\smC}{{\rm\scriptscriptstyle C}}
\newcommand{\smR}{{\rm\scriptscriptstyle R}}
\newcommand{\smT}{{\rm\scriptscriptstyle T}}
\newcommand{\smS}{{\rm\scriptscriptstyle S}}

\newcommand{\smLT}{{\rm\scriptscriptstyle <}}

\newcommand{\smD}{{\rm\scriptscriptstyle D}}

\begin{document}

\preprint{LA-UR 11-03256}

\title{
Charged Particle Motion in a Plasma: \\
Electron-Ion Energy Partition
}

\author{Lowell~S. Brown, Dean~L. Preston, and Robert~L. Singleton Jr.}

\affiliation{
     Los Alamos National Laboratory\\
     Los Alamos, New Mexico 87545, USA
}
\date{\mydates}

\begin{abstract}
\abstractskip

This paper considers plasmas in which the electrons and ions may have
different temperatures. This is a case that must be examined because 
nuclear fusion processes, such as those that appear in ICF capsules, 
have ions whose temperature runs away from the electron temperature.
A fast charged particle traversing a plasma loses its energy to both
the electrons and the ions in the plasma.  We compute the energy
partition, the fractions $E_e / E_0 $ and $ E_\smI / E_0$ of the
initial energy $E_0$ of this `impurity particle' that are deposited
into the electrons and ions when it has slowed down into a
``schizophrenic'' final ensemble of slowed particles that has neither
the electron nor the ion temperature. This is not a simple Maxwell-Boltzmann
distribution since the background particles are not in thermal equilibrium.
We perform our calculations using a 
well-defined Fokker-Planck equation for the phase space distribution
of the charged impurity particles in a weakly to moderately coupled
plasma.  The Fokker-Planck equation holds to first sub-leading order
in the dimensionless plasma coupling constant, which translates to
computing to order $n\ln n$ (leading) and $n$ (sub-leading) in the
plasma density $n$.  An examination of the energy partition for the 
general case, in which the background plasma contains two different 
species of particles that are not in thermal equilibrium, has not been
previously presented in the literature. We have new results for this case. 
The energy partitions for a background plasma in thermal equilibrium
have been previously computed, but the order $n$ terms have not been
calculated, only estimated.  Since the charged particle does not come
to rest, but rather comes into a statistical distribution, the energy
loss obtained by a simple integration of a $dE/dx$ has an ambiguity on
the order of the plasma temperature.  Our Fokker-Planck formulation
provides an unambiguous, precise definition of the energy
fractions. For equal electron and ion temperatures, we find that our
precise results agree well with a fit obtained by Fraley, Linnebur,
Mason, and Morse.   The ``schizophrenic'' final ensemble of slowed
particles gives a new mechanism to bring the electron and ion
temperatures together. The rate at which this new mechanism brings the
electrons and ions in the plasma into thermal equilibrium will be
computed.

\end{abstract}

\maketitle
\bodyskip

\section{Introduction}

The underlying theme of this paper is the thermonuclear burn of
deuterium-tritium plasmas. We do not consider the initiation of the
burn process, which is system specific, nor are we interested in the
late stages of the process when most of the DT fuel has been burned
into alpha particles and neutrons, and the electrons and ions are
nearly in thermal equilibrium. We instead focus on intermediate times
when, in general, there is a significant difference between the
electron and ion temperatures, but the alpha particle density has not
yet become a significant fraction of the D and T ion
densities.\footnote{When the alpha particle density is a significant
  fraction of the plasma ion density, the effect of the alphas on the
  dielectric response of the plasma must be taken into account. This
  introduces additional complications, and as such merits a separate
  publication.}  The fusion rate is very sensitive to the ion
temperature $T_\smI$. The ion temperature is determined by competition
between deposition of the alpha particle energy into the ions, which
of course increases $T_\smI$, and thermal equilibration with the
electron distribution, which drives $T_\smI$ down. Our main concern in
this paper is the partition of the total alpha energy between the ions
and electrons in a two-temperature plasma in the circumstances that we
have outlined.\footnote{A short preliminary account of the methods
  that we employ in this paper, but restricted to the case of equal
  ion and electron temperatures, has previously been presented in
  \cite{SB}.} This is important in the understanding of the time scale
and the robustness of the fusion process.  Our evaluations of the
functions which determine the energy partition do not include a
contribution from the alpha particles; hence our results are valid
only if the ensemble of alphas is sufficiently dilute. We find that
the alpha particles slow down into a non-Maxwellian distribution in
which the mean alpha energy $\bar E$ lies between the thermal energies
of the ions and electrons. Our work shows that these non-thermal alpha
particles increase the rate of energy transfer between the electrons
and ions but, since we do not examine late times where the population
of alpha particles is large, this new mechanism does not significantly
enhance the energy transfer rate. In general, as in other work on
stopping power and the partition of a fast impurity particle's energy
to the electrons and ions in the plasma, we assume (as is most often
the case) that the stopping times are much shorter than the time scale
of the fusion so that we can work in the adiabatic approximation in
which the time dependences of our results are only those brought about
by the changes in the plasma parameters on which they depend.  We also
require, as is also generally assumed, that the charged particle range
is short in comparison with the distances over which the plasma
conditions vary so that the plasma may be treated as being uniform.

The major results of this paper are as follows. First, as we have
mentioned in the previous paragraph, we have worked out the energy
partition for differing electron and ion temperatures; this has not
been previously considered in the literature. Second, even for the
case of equal ion and electron temperatures, where the alphas relax
into a Maxwellian distribution, we have made two improvements. We have
developed a formulation that precisely defines the energy partition so
that a correction of order $T / E_0$ is now included, a correction
that is missing in the literature. In addition to the well-known $n\,
\ln\, n$ ($n$ is the number density) terms in the energy partition, we
have computed exactly the coefficient of the order $n$ term, which has
previously been only estimated. We turn now to describe our work in
some detail.
 
When a fast charged particle with initial energy $E_0$ traverses a
plasma, it loses its energy at a rate $dE/dx$ per unit of distance,
and it comes into a quasi-static equilibrium state after depositing
its initial energy into the electrons and ions that make up the
plasma. In the thermonuclear fusion process of deuterium and tritium,
$D + T \to n + \alpha$, which occurs in inertial confinement fusion
experiments, the amount of the initial alpha-particle energy
$E_0 = 3.54$~MeV that is transferred to the $D,T$ ions is
crucial because a high ion temperature is necessary for the fusion
reaction parameter $\langle \sigma \, v \rangle_\smT$ to become
sufficiently large so as to have a robust and stable fusion burn.

In the picture in which the projectile traverses linearly through the
plasma until coming to a complete stop, the energy partition into ions
and electrons is given by
\begin{equation}
  E_\smI 
  = 
  \int_0^{E_\smI}\! dE_\smI
  = 
  \int_0^{E_0}\! dE ~ \frac{dE_\smI/dx}{dE/dx} \,,
\label{eion}
\end{equation}
and  
\begin{equation}
  E_e 
  = 
  \int_0^{E_e}\! dE_e
  = 
  \int_0^{E_0}\! dE ~ \frac{dE_e/dx}{dE/dx} \,.
\label{eelectron}  
\end{equation}
Here $dE_\smI/dx$ and $dE_e/dx$ are the stopping power contributions
from the ions and electrons, and $dE/dx$ is the total stopping power,
\begin{equation}
  \frac{dE}{dx} = \frac{dE_\smI}{dx} + \frac{dE_e}{dx} \,,
\end{equation}
and thus
\begin{equation}
  E_\smI + E_e = E_0 \,.
\end{equation}
This simple picture, however, is only an approximation. For a plasma
with equal ion and electron temperatures, a
fast charged particle does not simply come to rest in the plasma, but
rather, it becomes thermalized at the ambient plasma temperature
$T= T_\smI = T_e$. Expressing temperature in energy units, as we shall do
throughout this paper, the correct electron-ion energy partition
relation should read
\begin{equation}
  E_\smI + E_e + \frac{3}{2}\,T= E_0 \,.
\label{EieT0}
\end{equation}
Consequently, rather than extending the lower limits of the integrals
(\ref{eion}) and (\ref{eelectron}) down to zero energy,  lower limits 
of order the temperature, $E_\text{min} \sim T$, must be chosen.
The two integrals (\ref{eion}) and (\ref{eelectron})  have somewhat 
different thermal cutoffs, both of order $T$, and this simple picture has
a systematic error of relative order $T/E_0$. We see that the correction
becomes more important as the plasma temperature is elevated. To
account for the energy partition in a precise fashion, we shall employ
the Fokker-Planck equation in the version introduced by Brown,
Preston, and Singleton (BPS) \cite{bps}.  We shall find that the
correct expression for the energy partition does not, in fact, involve
the stopping powers $dE_\smI/dx$ and $dE_e/dx$, but rather certain ion 
and electron functions ${\cal A}_\smI$ and ${\cal A}_e$ that enter into 
this Fokker-Planck equation. In the notation of BPS, the stopping power 
of a particle of energy $E = \frac{1}{2} \,m \, v^2$ is of the generic 
form
\begin{equation}
  \frac{dE_a}{dx} 
  =  
  \left[ 1 - \frac{T}{m v} \, \frac{\partial}{\partial {\bf v}} \,
 \cdot  \hat {\bf v} \right] \,  {\cal A}_a \,.
\label{stopping}
\end{equation}
The functions ${\cal A}_\smI$ and ${\cal A}_e$ thus approach
$dE_\smI/dx$ and $dE_e/dx$ at high energies, but differ from these
stopping powers at low energies on the order of the thermal 
background temperature.

For the case of equal electron and ion temperatures, the explicit 
evaluations for $E_\smI$ and $E_e$ derived in Eqs.~(\ref{IIIanswer}) 
and (\ref{eeeanswer}), which omit of negligibly small
exponential terms involving $\exp\{ - \beta E_0\}$, read:
\begin{eqnarray}
  E_\smI 
  &=& 
  \int_0^{E_0} dE \,\, \frac{ {\cal A}_\smI(E)}{{\cal A}(E) }
  \left[ {\rm erf}(\sqrt{\beta E} ) -  
  \sqrt{ \frac{ 4\beta  E }{\pi} }  \,  e^{-\beta E} \right] \,,
\label{IIIIanswer}
\end{eqnarray}
and
\begin{eqnarray}
  E_e 
  &=& 
  \int_0^{E_0} dE \,\, \frac{ {\cal A}_e(E)}{{\cal A}(E) }
  \left[ {\rm erf}(\sqrt{\beta E} ) -  
  \sqrt{ \frac{ 4\beta  E }{\pi} }  \,  e^{-\beta E} \right] \,,
\label{eeeeanswer}
\end{eqnarray}
where
\begin{equation}
{\cal A}(E) = {\cal A}_\smI(E) + {\cal A}_e(E) \,,
\end{equation}
and $\beta = 1 /T$. Here ${\rm erf}(x)$ is the error function defined
in Eq.~(\ref{erf}). Using the definition (\ref{erf}), partial
integration can then be used to show that the sum rule (\ref{EieT0})
follows from Eqs.~(\ref{IIIIanswer}) and (\ref{eeeeanswer}).  Since
$dE_b/dx \to {\cal A}_b$ for large energies, and since the error
function approaches unity at large $\beta E$, the precise results
(\ref{IIIIanswer}) and (\ref{eeeeanswer}) approach the more intuitive
but less accurate forms (\ref{eion}) and (\ref{eelectron}).
Significant differences occur only for $E \sim T$.

\begin{figure}[h!]
\vskip-1.5cm
\includegraphics[scale=0.50]{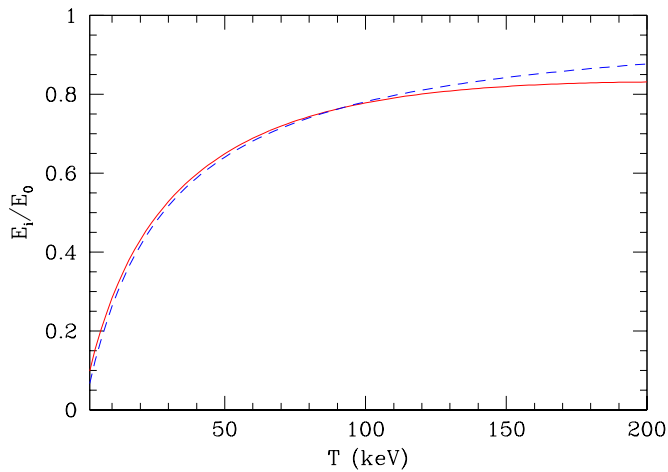}
\vskip-1cm 
\caption{\captionskip
  The fractional energy loss into ions as a function of the plasma
  temperature for an $\alpha$ particle in an equimolar DT plasma with
  initial energy $E_0=3.54\,{\rm MeV}$. The electrons and ions have a 
  common temperature $T$ and the electron number density of the plasma
  is $n_e=1.0 \times 10^{26}\,{\rm cm}^{-3}$.  The solid red line is 
  the evaluation of Eq.~(\ref{IIIIanswer}) while the dashed blue line 
  is the FLMM fit (\ref{rule}) with $T_\smC=28\,{\rm keV}$ rather than 
  the $32\,{\rm keV}$ value used by FLMM. 
}
\label{fig:frac26}
\end{figure}

\begin{figure}[ht!]
\vskip-2.6cm
\includegraphics[scale=0.50]{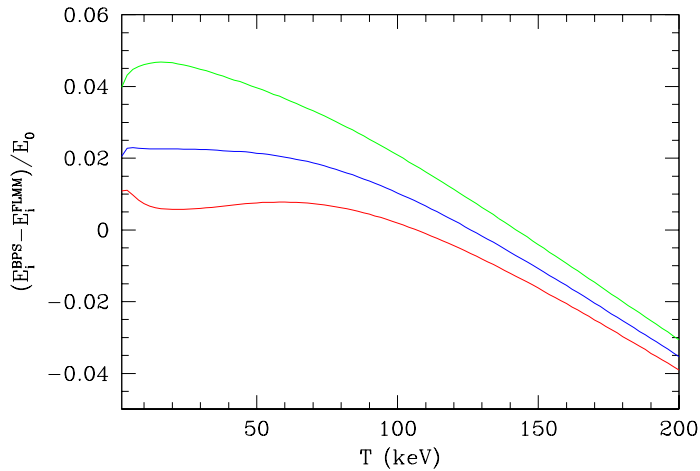}
\vskip-1cm 
\caption{\captionskip
  Differences between the fractional energy losses $E_\smI/E_0$ as 
  given by the precise result (\ref{IIIIanswer}) and the FLMM fit 
  (\ref{rule}) for an alpha particle with initial energy $E_0= 3.54$ MeV 
  in an equimolar DT plasma.  The different curves correspond to the 
  three electron densities $n_e$ of $1.0 \times 10^{24} \, {\rm cm}^{-3}$ 
  (red), $1.0 \times  10^{25} \, {\rm cm}^{-3}$ (blue), 
  and $1.0 \times 10^{26} \, {\rm cm}^{-3}$ (green), with the fit 
  (\ref{rule}) evaluated with 
  $T_\smC=31 \,,\, 30\,,\, {\rm and} \,\, 28 \, {\rm keV}$,  
  respectively.  }
\label{fig:change}
\end{figure}

We have numerically evaluated these integrals using the expressions for 
the ${\cal A}$ functions derived in BPS that are reproduced in 
Appendix \ref{app:acoeff}. We shall compare our results with the 
less precise but well known results of Fraley, Linnebur, Mason, and Morse 
\,(FLMM)\,\cite{fraley}. Starting with a model of the stopping power in 
an equimolar DT plasma, these authors show that the simple rule
\begin{equation}
  \frac{E_\smI}{E_0} = \frac{1}{1 + T_\smC/ T_e}  
\label{rule}
\end{equation}
provides a good fit to their calculations. The crossover temperature
$T_\smC$, where the electron and ion fractions are equal, can be
determined from their Fig.~1b. Fraley {\em et al.} find
$T_\smC=32\,{\rm keV}$ at the density $\rho=0.213\,{\rm g/cm^3}$,
or a corresponding electron number density $n_e= 5.0 \times
10^{22}\,{\rm cm}^{-3}$. At the number densities $n_e= 1.0 \times
10^{24}
 \,,\, 1.0 \times 10^{25} \,,\, 1.0 \times 10^{26}\,{\rm cm}^{-3}$, 
we find, by fitting our more precise results, 
 that $T_\smC=31 \,,\, 30 \,,\, 28 \,\, {\rm keV}$ respectively.
Figure~\ref{fig:frac26} shows our result (\ref{IIIIanswer}) for the 
fractional energy loss to ions and the FLMM fit (\ref{rule}) for a DT 
plasma with electron number density $n_e=10^{26}\,{\rm cm}^{-3}$. In 
this comparison, we use the more accurate value $T_\smC = 28$ keV.
In Fig.~\ref{fig:change} we compare the differences between our result 
(\ref{IIIIanswer}) and the FLMM fit (\ref{rule}) over a wide range of 
densities.  We see
that the FLMM fit somewhat overestimates the energy deposited to ions for
temperatures above 120\,{\rm keV} over a wide range of densities.
\begin{figure}[ht!]
\includegraphics[scale=0.45]{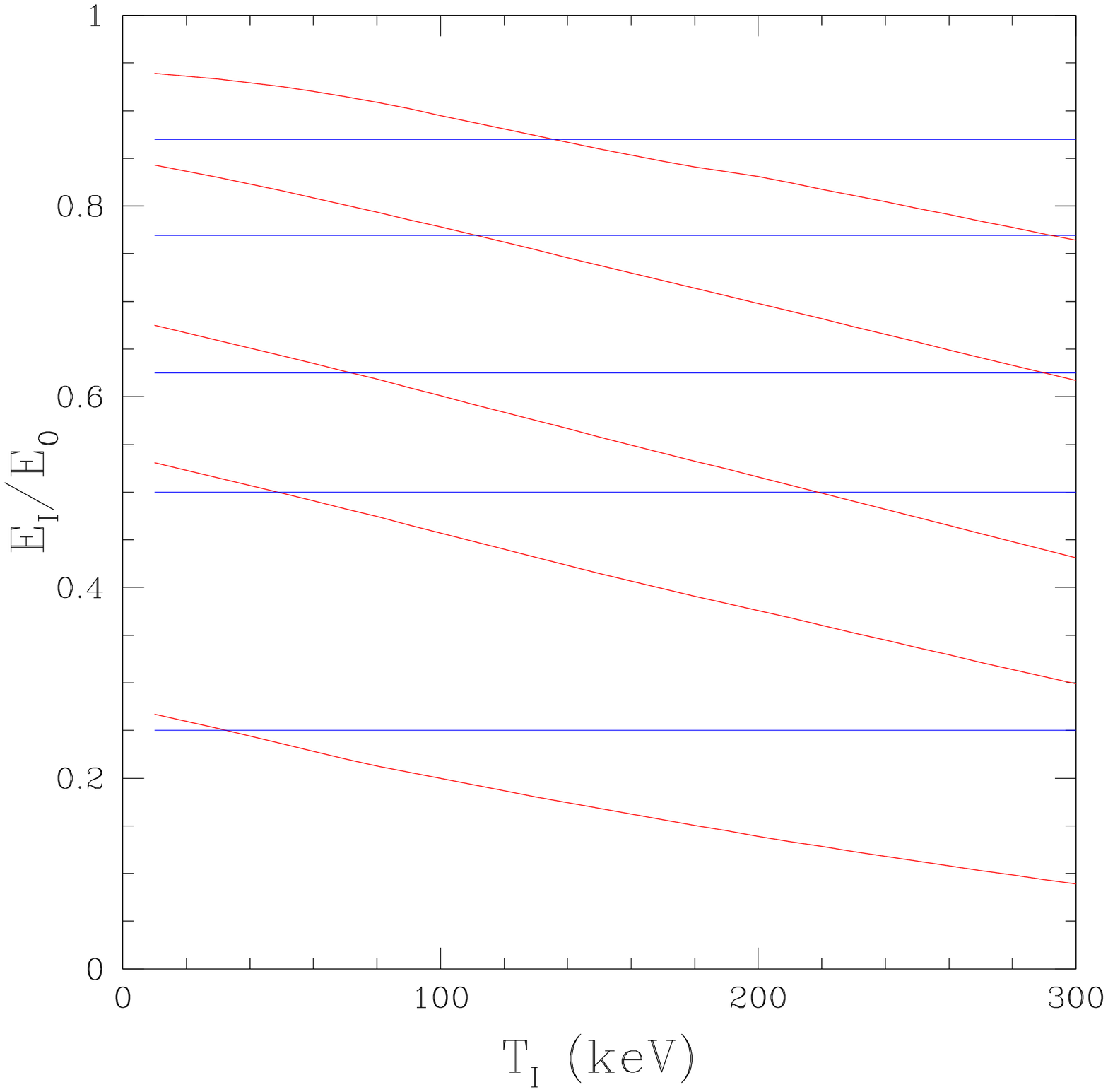}
\vskip-0.2cm 
\caption{\footnoteskip  
  The red curves show the energy fractions deposited to the ions as 
  functions of the ion temperature $T_\smI$ for various values of the 
  electron temperature $T_e$ for an $\alpha$ particle with initial 
  energy $E_0=3.54 \, {\rm MeV}$ in an equimolar DT plasma with an 
  electron density $n_e = 1.0 \times 10^{25} \, {\rm cm}^{-3}$.  The blue
  horizontal lines are the energy fractions determined by the FLMM 
  fit (\ref{rule}) with our value $T_\smC = 30 \,{\rm keV}$. 
  The red curves describing larger values of $E_\smI / E_0 $ correspond to 
  increasing electron temperatures of 
  $T_e = 10 \,,\, 30 \,,\, 50 \,,\, 100 \,,\, 200 \, {\rm keV}$.
  If the fit (\ref{rule}) were exact, the red curves would cross
  the blue horizontal lines when $T_\smI = T_e$. This condition is
  fairly well met except at the highest electron temperature 
  $T_e = 200$ keV where the value of $E_\smI / E_0$ given by the 
  red curve at $T_\smI = 200$ keV is considerably smaller than the 
  blue horizontal line.  This is consistent with the discrepancy at
  these temperatures shown in Fig.~\ref{fig:frac26}. }
\label{fig:TiEbarFig3.eps} 
\end{figure}

As Figs.~\ref{fig:frac26} and \ref{fig:change} show, the FLMM fit
(\ref{rule}), modified slightly to use better values of the crossover
temperature $T_\smC$, is in good agreement with our precise results in
the case of equal temperatures so long as these temperatures are less
that about $120$ keV.  However, as Fig.~\ref{fig:TiEbarFig3.eps}
demonstrates, this simple form fails to provide an accurate estimate
of the energy partition when the ion and electron temperatures are
significantly different.  These results for differing electron and 
ion temperatures follow from Eqs.~(\ref{IIIIfrac}) and (\ref{done}).
They are spelled out in more detail in the tables presented in 
our concluding section \ref{Con}. 

Although the work of FLMM continues to be used, a more recent
evaluation of the energy partition has been carried out by Li and
Petrasso \cite{LP}.  Comparing their Table 1 ($T_\smI = T_e$) with our
Fig.~\ref{fig:frac26} shows that their results for $E_\smI / E_0$ are
$\sim 10\%$ too high.\footnote{A detailed discussion 
of the results of Li and Petrasso \cite{LP} for the stopping power
$dE/dx$ was presented in the BPS paper \cite{bps} that provides the
basis for the work which we perform here.}  This discrepancy 
is of order the sub-leading corrections to the Coulomb 
logarithm.\footnote{Long and Tahir
\cite{LT} have also presented results for the energy partition, but
they only compute the separate electronic and ionic contributions to
the stopping powers, $dE_e / dx$ and $dE_\smI/dx$, as a function of
the range $x$ for equal temperature background plasmas. They do not
present the total energies deposited to the electrons and ions, and
they also do not present a precise Coulomb logarithm.}

The emphasis in this paper is the energy partition for unequal
electron and ion temperatures, which is of the utmost importance for
DT burn since there the ion temperature generally runs away from the
electron temperature once the fusion process begins. To describe this
in a simple fashion, we assume that the sources of the charged
impurity particles (the $\alpha$ particles in DT fusion) are uniformly
distributed throughout the plasma, and that the particles are emitted
isotropically; hence, the phase space distribution of the impurity
particles is only a function of the energy and time. The evolution of
this distribution is governed by a Fokker-Planck equation that
involves the coefficient functions ${\cal A}_\smI(E)$ and 
${\cal A}_e(E)$ which were computed in BPS to order $ n (\ln n +c) $ in 
the plasma density $n$. Since $n \sim g^2$, with $g$ the plasma coupling
constant, it is evident that these two terms in the density are the
leading and first sub-leading terms in the perturbative expansions in
$g$ of the coefficient functions. Higher-order terms in the expansions
become significant at high densities, hence our results are not
applicable, in particular, to (strongly coupled) warm dense plasmas.
Numerical simulations provide the only potentially reliable means of
validating our analytic expressions for the energy partition in weakly
coupled plasmas and evaluating the partition in moderately to strongly
coupled plasmas, though such computations have not been performed.
Careful, large statistics, molecular dynamics (MD) simulations have
been carried out by Dimonte and Daligault \cite{DD} to investigate 
electron-ion temperature relaxation over a wide range of plasma
parameters that span weak to strong coupling. Their
MD results for the Coulomb logarithm for this process agree with those
of BPS \cite{bps} for $g < 0.2$ to within the statistical uncertainty of $\pm
5\%$ in the simulations. This indicates the range of validity of the
Fokker-Planck equation that we use to compute a different result, 
the energy partition.

Following a detailed discussion of the Fokker-Planck equation in
Sections \ref{Fokk} and \ref{longit}, the late-time distribution
$f_\infty(E)$ of a $\delta(t)$ source of impurity particles, which is
needed to obtain the electron-ion energy split, is derived in Section
\ref{asymsol}. In Section \ref{sec:evolution} a source is slowly
turned on and eventually emits particles at a constant rate. The
solution $f(E,t)$ of the now inhomogeneous Fokker-Planck equation is
shown to be the sum of two terms: $ f(E,t) = n(t) \, f_\infty (E) +
\bar f(E)$, where $n(t)$ is the number density of impurity particles
that have come into the equilibrium state described by $f_\infty(E)$,
and $\bar f(E)$, which describes the transfer of energy to the
electrons and ions. The energy losses $E_e$ and $E_\smI$ to the
electrons and ions are expressed as single integrals involving the
function $\bar f(E)$ [which depends upon the ${\cal A}$-coefficients]
and the ${\cal A}$-coefficients themselves.  The late-time ensemble of
impurity particles with energy distribution $f_\infty(E)$ is not in
thermal equilibrium with the background plasma, {\it i.e.}
$f_\infty(E)$ is not a Maxwell-Boltzmann distribution. This ensemble
increases the rate of ion-electron thermal equilibration above that of
the impurity-free plasma. 
In Section \ref{gendev} we carry out the explicit construction of
$\bar f(E)$. We show how our general results for the deposited energy
fractions $E_\smI$ and $E_e$ in the equal temperature case reduce to
the usual expressions involving $dE_\smI/dx$ and $dE_e/dx$ in Section
\ref{EfracdE/dx} and then describe how these approximate results are
corrected with our precise formulation in Section \ref{equaltemps}.
In Section \ref{differing} we compute $E_\smI/E_0$ and $E_e/E_0$
for the general case of different plasma electron and ion temperatures
in terms of integrals over ${\cal A}_\smI$ and ${\cal A}_e$.
The conclusion \ref{Con} provides a summary of our major results
including a table of the energy fractions $E_\smI/E_0$ and $E_e/E_0$
for a wide range of plasma parameters. 
At this point, we have finished a logically complete exposition of
our methodology and results, which is essentially
self-contained. However, for those interested in supporting details
and who may wish to work out the intermediate steps in our
calculations, we include these details in the Appendices.
We provide a review of the ${\cal A}$ functions that
were computed in BPS \cite{bps} which are needed for the present work
in Appendix \ref{app:acoeff}, a host of details on these functions that
include their approximate forms in various regions in Appendix
\ref{sec:limits}, and an accurate approximation for one of the two
multiple integrals appearing in our final expressions for $E_\smI$ and
$E_e$  is provided in Appendix \ref{hack}.

\section{Formulation of the Problem}

\subsection{The Fokker-Planck Equation to Leading and Next-to-Leading Order}
\label{Fokk}

We consider a plasma containing a dilute population of ``impurity''
particles with a phase space density $f({\bf r}, {\bf p}, t)$.  For
example, in a deuterium-tritium (DT) plasma, the impurities could
consist of the charged $\alpha$ particles produced from the DT fusion.
The problem we shall address is the manner by which such impurities
reach a quasi-static equilibrium distribution. During this
process, the impurities deposit portions of their energy to plasma
electrons and plasma ions, and the formalism we now develop will allow
us to compute the electron-ion energy splitting in a systematic and
unambiguous fashion. We take the plasma to have an electron
temperature $T_e = \beta_e^{-1}$ and a common temperature
$T_\smI = \beta_\smI^{-1}$ for all the ions, in which case the
Fokker-Planck equation for the distribution $f$ of an impurity species
has the form
\begin{equation}
  \left[ \frac{\partial}{\partial t} + {\bf v} \cdot 
  {\bm\nabla} \right]
   f({\bf r},{\bf p},t)  
  = {\sum}_b \, 
  \frac{\partial}{\partial p^k} \, C_b^{k\ell}({\bf p}) 
  \left[\beta_b v^\ell + \frac{\partial}{\partial p^\ell} \right] 
  f({\bf r},{\bf p},t) \,,
\label{FP}
\end{equation}
where ${\bf v} = {\bf p} / m $ is the velocity of an impurity particle
with momentum ${\bf p}$, the explicit sum runs over all the particle
species $b$ in the background plasma, and the summation convention is
used for repeated vector indices $k$ and $\ell$.  As we shall describe
more fully, the diffusion coefficient $C_b^{k\ell}$ has been 
analytically calculated to
leading and next-to-leading orders in the plasma density in
BPS~\cite{bps} or more precisely, to orders $g^2 \ln g^2$ and
$g^2$ in the generic dimensionless plasma coupling constant
$g = e^2 \kappa/4 \pi T$. We use rationalized
electrostatic units, so that this parameter is the Coulomb
energy of two particles of charge $e$ a Debye distance $1 / \kappa$ 
apart divided by an average temperature  $T$. 

With our conventions, the number of impurity particles is given by
\begin{equation}
  N(t) 
  = 
  \int d^3 r \int \frac{d^3 p}{(2\pi\hbar)^3} \,  
  f({\bf r},{\bf p},t) \,,
\label{number}
\end{equation}
and their kinetic energy and momentum appear as
\begin{equation}
  E(t) = 
  \int d^3 r \int \frac{d^3 p}{(2\pi\hbar)^3} \,
  \frac{p^2}{2m} \, f({\bf r},{\bf p},t) \,,
\label{energy}
\end{equation}
and
\begin{equation}
  {\bf P}(t) = 
  \int d^3 r \int \frac{d^3 p}{(2\pi\hbar)^3} \,
  {\bf p} \, f({\bf r},{\bf p},t) \,.
\label{momentum}
\end{equation}
Since the right-hand side of the Fokker-Planck equation (\ref{FP})
contains an overall total momentum derivative, it
does not contribute to the time rate of change of the particle number
--- the Coulomb collisions in the plasma preserve particle number.
When the electrons and ions are at common temperature $T =
\beta^{-1}$, the terms in the final square brackets in the
Fokker-Planck equation (\ref{FP}) annihilate a thermal
Maxwell-Boltzmann distribution [$ f \propto \exp\{ - \beta\, {\bf p}^2
/ 2m \} $] of impurity particles --- a collection of particles in thermal
equilibrium is not altered by their collisions with a background plasma
at the same temperature.  However, for those cases
in which the ions and electrons have different
temperatures, the ``injected impurity particles'' attain a non-thermal
quasi-static distribution that will be described shortly. Eventually
this quasi-static distribution will relax into a thermal distribution
as the electron and ion components themselves thermally relax. As we
shall see, however, the impurity distribution has interesting effects
on temperature relaxation at intermediate times.

The stopping power can be extracted from the Fokker-Planck
equation by considering a single impurity particle at ${\bf r}_p$ 
moving with the velocity ${\bf  v}_p$.  The corresponding 
distribution function is given by 
$ f_p({\bf  r},{\bf p},t) = (2\pi\hbar)^3 \delta({\bf r} - {\bf r}_p)
\delta({\bf p} - {\bf p}_p) , $ 
and one can easily check that this
distribution indeed gives $N=1$ as it should.  Inserting this
single particle distribution into Eq.~(\ref{FP}) and performing a
partial integration, it is easy to see that 
the rate of energy loss of the particle is given by
\begin{equation}
  \frac{dE}{dt} 
  =  +  {\sum}_b \,
  \left[ \beta_b v^\ell_p - \frac{\partial}{\partial p^\ell_p} \right] \, 
  v^k_p \, C_b^{k\ell}({\bf p}_p) \,.
\label{dedtckl}
\end{equation}
To make the sign of this expression clear, we emphasize that it
gives the rate at which the particle  {\em loses}
energy to the plasma  [it is the negative of the time derivative
of Eq.~(\ref{energy})]. Hence the stopping power, which is the
energy loss of the particle per unit distance traveled, appears as
\begin{equation}
  \frac{dE}{dx} =  +\, \frac{1}{v_p} \, \frac{dE}{dt}  \,.
\end{equation}
In a similar manner, we can find the rate of change of the momentum
by substituting the single particle distribution into expression
(\ref{momentum}), thereby giving 
\begin{equation}
  \frac{dP^k}{dt} 
  =  
  {\sum}_b \,\left[ \beta_b v^\ell_p - \frac{\partial}{\partial p^\ell_p} 
  \right] \, C_b^{k\ell}({\bf p}_p) \,.
\label{dpdtckl}
\end{equation}
As performed in BPS, by calculating $dE/dt$ and $dP^k/dt$ to leading
and next-to-leading order, we can invert equations (\ref{dedtckl}) and
(\ref{dpdtckl}) to the same order to obtain the coefficients $C_b^{k\ell}$
of the Fokker-Planck equation. 

\subsection{Longitudinal and Transverse Components of the Diffusion Tensor}
\label{longit}

As described in detail in BPS, the isotropy of the background thermal 
plasma allows one to decompose the diffusion tensor as
\begin{equation}
  C_b^{k\ell}({\bf p}) 
  = 
  {\cal A}_b(v) \, \frac{\hat v^k \hat v^\ell}{\beta_b v} +
  {\cal B}_b(v) \, \frac{1}{2} \, \left( \delta^{k\ell} - 
  \hat v^k \hat v^\ell \right) \,,
\label{CaklAB}
\end{equation}
where $v$ is the magnitude of the velocity,  $v = | {\bf v}|$,
with the velocity direction given by  $ \hat {\bf v} = {\bf v} / v $.
We often take the independent variable to be the
energy $E=\frac{1}{2}\,m v^2$ and, with a slight abuse of notation, we
shall also write ${\cal A}_b={\cal A}_b(E)$ and ${\cal B}_b={\cal
  B}_b(E)$.  As a matter of completeness, the ${\cal A}$-coefficients
are provided in Appendix~\ref{app:acoeff}, and their various
limits can be found in Appendix~\ref{sec:limits}.  For a
homogeneous and isotropic source of impurity particles, the case we
shall consider, the ${\cal B}$-coefficients do not enter, although
their analytic forms can be found in BPS~\cite{bps} if desired.

Let us return to the stopping power (\ref{dedtckl}) of a charged particle.
Since the velocity tensor multiplying the ${\cal B}$-contribution is
transverse --- its contraction with $v^k$ or $v^\ell$ vanishes --- the
rate of energy loss (\ref{dedtckl}) of a projectile becomes
\begin{equation}
  \frac{dE}{dt} =  {\sum}_b \,
  \left[ v - \frac{1}{\beta_b m} \, \frac{\partial}{\partial v^\ell} \,
  \hat v^\ell \right] \,  {\cal A}_b \,,
\label{Edott}
\end{equation}
where we have now omitted the $p$ subscript.  The respective energy
losses to the ions and electrons are given by separating this formula
into the ion contribution described by
\begin{equation}
  {\cal A}_\smI = {\sum}_i \, {\cal A}_i \,,
\end{equation}
and the electron part governed by ${\cal A}_e$, so that\footnote{
   As noted in BPS, to the order in $g$ 
   in which we are working, namely to leading ($g^2\ln g^2$) and 
   next-to-leading ($g^2$)  order, only the kinetic energy of the 
   stopping ion enters, and a meaningfully separation into electron 
   and ion energy components can be made. This is because of the 
   trivial fact that the kinetic energy is independent of $g$ --- 
   it is of order $g^0$. In addition to this kinetic energy, the
   impurity particle has potential energy interactions with the 
   ions in the background plasma. The change in these interaction 
   energies associated with the motion of an impurity particle in a 
   plasma cannot be separated into different parts that are 
   associated with the ions and with the electrons.  This is
   because this potential energy  starts out at order $g$, 
   and thus its evolution, which involves interactions akin to 
   those involved in the kinetic energy $dE/dx$, is of order 
   $g^3$ (modulo possible logarithms), an order that is higher 
   than that considered in this paper. Thus it should be 
   emphasized that at higher orders in $g$, such clean separation 
   into energies deposited into well-defined, separate ion and
   electrons components cannot be performed.
}
\begin{equation}
  \frac{dE_\smI}{dt} 
  =  
 \left[ v - \frac{1}{\beta_\smI m} \, \frac{\partial}{\partial v^\ell} \,
 \hat v^\ell \right] \,  {\cal A}_\smI 
\label{IEdott}
\end{equation}
and
\begin{equation}
  \frac{dE_e}{dt} 
  =  
  \left[ v - \frac{1}{\beta_e m} \, \frac{\partial}{\partial v^\ell} \,
  \hat v^\ell \right] \,  {\cal A}_e \,,
\label{eEdott}
\end{equation}
with their sum giving
\begin{equation}
  \frac{dE}{dt} =  \frac{dE_\smI}{dt} +  \frac{dE_e}{dt} \,.
\end{equation}
The rates $dE_b/dt$ were rigorously computed in BPS to the leading $g^2 \ln
g^2$ and sub-leading $g^2$ orders, and these results were then used to 
determine ${\cal A}_b$
to these orders. In a similar fashion, BPS also computed the rate of
momentum change $d{\bf P}_b/dt$ of a projectile to these orders to
determine the other coefficients ${\cal B}_b$.  In this way, a
Fokker-Planck equation was determined to these orders in an
unambiguous manner with no undetermined
parameters.\footnote{
  See BPS \cite{bps} for a full discussion of the range of validity of
  the Fokker-Planck equation constructed in this fashion.  }
In particular, we should emphasize that our Fokker-Planck
equation describes a particle's energy loss including orders $g^2
\ln g^2$ and $g^2$ with no ambiguity.

Rather than tracking an individual charged particle slowing
down in the plasma, it is much simpler --- and equivalent ---
to examine an isotropic distribution of particles. 
When the impurity distribution is  isotropic,
$f$ is a function the magnitude of the momentum 
$p=\vert {\bf p}\vert$ or equivalently, of the speed $v$ or
energy $E$. In such cases, a momentum derivative of $f$ produces
a factor of the velocity vector whose contraction with the
velocity tensor multiplying the ${\cal B}_b$ coefficients
vanishes. Hence in the isotropic case, the Fokker-Planck 
equation (\ref{FP}) reduces to
\begin{equation}
  \left\{ \frac{\partial}{\partial t} - 
  \frac{\partial}{\partial {\bf v}} \cdot \hat {\bf v} \,
  {\sum}_b \frac{{\cal A}_b}{m}\,\left[ 1 + 
  \frac{\hat {\bf v}}{\beta_b m v}\cdot \frac{\partial}{\partial{
  \bf v}} \right] \right\} \, f(E,t) 
  =  
  0 \ .
\label{fphomo}
\end{equation}
To avoid notational clutter, we  define the total 
${\cal A}$-coefficient by
\begin{equation}
  {\cal A}(E) = {\cal A}_\smI(E) + {\cal A}_e(E) \,,
\label{calEone}
\end{equation}
and the temperature-weighted ${\cal A}$-coefficient by
\begin{equation}
  \langle T{\cal A}(E) \rangle 
  = 
  T_\smI \, {\cal A}_\smI(E) +   T_e \, {\cal A}_e(E)  \,.
\label{calEtwo}
\end{equation}
Thus,
\begin{equation}
  \left\{ \frac{\partial}{\partial t} - 
  \frac{\partial}{\partial {\bf v}} \cdot \hat {\bf v} \,
  \left[ \frac{ {\cal A}(E) }{m}  + 
  \frac{  \langle T{\cal A}(E) \rangle }{m^2 v}
 \, \hat {\bf v} \cdot \frac{\partial}{\partial{
  \bf v}} \right] \right\} \, f(E,t) 
  =    0 \,.
\label{ffphomo}
\end{equation}
Using the operator forms
\begin{equation}
 \frac{\partial}{\partial{\bf v}} \cdot 
  \hat {\bf v} 
  = 
  v^{-2} \, \frac{\partial}{\partial v} \, v^2 
  = \frac{2}{v} \, \frac{\partial}{\partial E} \, E \,,
\label{usefulEQ}
\end{equation}
and
\begin{equation}
  \hat{\bf v} \cdot \frac{\partial}{\partial {\bf v}}
  = \frac{\partial}{\partial v} = 
  m v \, \frac{\partial}{\partial E} \,,
\end{equation}
we may express Eq.~(\ref{ffphomo}) in the form
\begin{equation}
  \left\{\frac{\partial}{\partial t} 
  - 
  \frac{1}{m v^2}\, \frac{\partial}{\partial v}\,
  \left[ v^2 \,{\cal A}(E)+ 
  \langle T{\cal A}(E)\rangle\,\frac{v}{m}\,
  \frac{\partial}{\partial v}
  \right] \right\} \, f(E,t) 
  = 0 \,,
\label{fphomov}
\end{equation}
or
\begin{equation}
  \left\{\frac{\partial}{\partial t} 
  - 
  \frac{2}{m v}\, \frac{\partial}{\partial E}\,
  E \left[ {\cal A}(E) + \langle T {\cal A}(E)\rangle\,
  \frac{\partial}{\partial E}
  \right] \right\} \, f(E,t) 
  = 0 \ .
\label{fphomoE}
\end{equation}

\subsection{Asymptotic Solution}
\label{asymsol}

As we shall see, to use these results to obtain an unambiguous
formulation of the fractions of the total energy deposited into the ions and
electrons, we first need to compute the asymptotic distribution into
which an initial swarm of test particles relaxes in the presence of a
background plasma of differing electron and ion temperatures. This
quasi-static distribution will be a function of $E$ (or equivalently of $p$),
which we express in terms of a function $S(E)$ as
\begin{equation}
  f_\infty(E) = {\cal N} \, e^{-S(E)} \,,
\label{finfty}
\end{equation}
where we choose ${\cal N}$ to normalize the distribution to unity, 
\begin{equation}
  1 = {\cal N} \,  \int\!\frac{d^3p}{(2\pi\hbar)^3} \, e^{-S(E)} \,.
\label{onenorm}
\end{equation}
The function $S(E)$ is determined by inserting the structure
(\ref{finfty}) into Eq.~(\ref{fphomoE}) which gives
\begin{equation}
  \frac{d}{d E} \, E \left\{  {\cal A}(E) + 
  \langle T{\cal A}(E) \rangle \, \frac{d}{d E} \right\} 
  e^{-S(E)} = 0 \,.
\label{homo}
\end{equation}

One solution of the second-order differential equation (\ref{homo}) is
obtained by requiring that the quantity in curly braces operating on
$\exp\{ - S(E) \} $ vanishes: 
\begin{equation}
  {\cal A}(E) - \langle T{\cal A}(E)\rangle\,\frac{d S(E)}{d E} = 0 \
  ,
\label{Seq}
\end{equation}
and the solution can be obtained by a simple integration
\begin{eqnarray}
  S(E; T_e, T_\smI) 
  &=& 
  \int_0^E dE'\,\frac{{\cal A}(E')}{\langle T{\cal A}(E') \rangle} \,.
\label{Sdef}
\end{eqnarray}  
Here we have temporarily indicated the explicate dependence
upon the electron and ion temperatures to emphasize that when the 
ions and electrons are at a common temperature $T = T_\smI = T_e$,
this solution reduces to the Maxwell-Boltzmann distribution
\begin{eqnarray}
  S(E; T, T) &=&  \frac{E}{T} \,,
\label{SdefT}
\end{eqnarray}  
and consequently a swarm of test particles simply relaxes to 
the background plasma equilibrium distribution.  
For the equal temperature solution (\ref{SdefT}), a simple analytic 
Gaussian  integration evaluates the normalization factor defined
in Eq.~(\ref{onenorm})  as
\begin{equation}
{\cal N} = \left( \frac{2\pi \hbar^2}{m T} \right)^{3/2} \,.
\end{equation}

Expression (\ref{Sdef}) is indeed the physical solution for $S(E)$.
This is because having the solution (\ref{Sdef}) in hand, it is a
matter of simple quadratures to construct the second,
linearly-independent solution for our second-order differential
equation (\ref{homo}).  It is not difficult to then confirm that this
second solution is not normalizable, and so our first solution is the
only physically relevant solution. We can also see that this is the
desired solution since, for equal temperatures, it relaxes to a
thermal Maxwellian distribution

\begin{figure}[t!]
\includegraphics[scale=0.45]{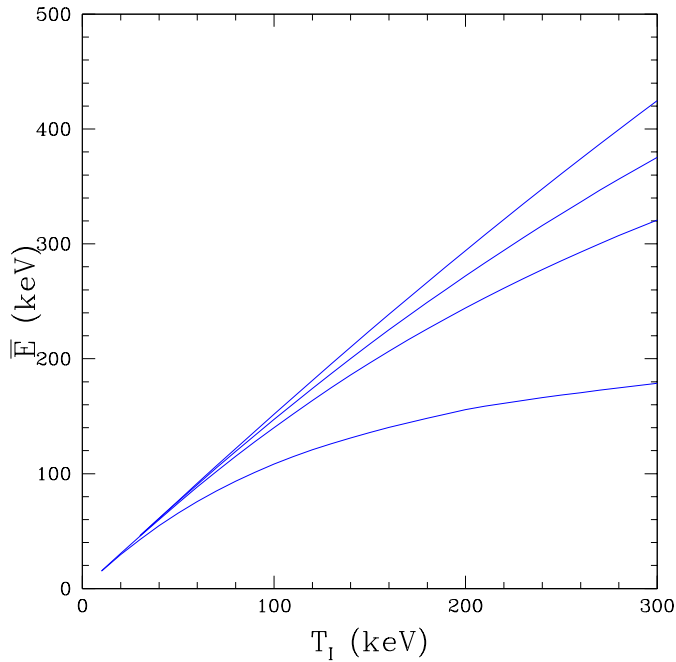}
\vskip-0.2cm 
\caption{\footnoteskip  
  Average energy $\bar E$ to which the $\alpha$ particle relaxes as a function
  of the ion temperature $T_\smI$ for various electron temperatures
  $T_e$. The ascending curves describing larger values of $\bar E$ have the 
  increasing electron temperatures 
  $T_e =  10 \,,\, 30 \,,\, 50 \,,\, 100$ keV.  
  When $T_e=T_\smI=T$ then $\bar E = \frac{3}{2}\,T$. The background plasma 
  is equimolar DT with electron number density 
  $n_e = 1.0 \times 10^{25}\,{\rm cm^{-3}}$. 
}
\label{fig:Fig1.eps} 
\end{figure}

The Maxwell-Boltzmann distribution has an average energy of
$ 3 T / 2$.  However, for the ions and electrons at different
temperatures, the swarm of test particles relaxes to the average
energy
\begin{equation}
  \bar E 
  = 
  {\cal N} \int\frac{d^3{\bf p}}{(2\pi\hbar)^3} \, 
  \frac{p^2}{2m}\,\exp\!\left\{-S\left( \frac{p^2}{2m}\right)\right\} \,.
\label{Ebar}
\end{equation}
In this case, numerical integrations are needed to
evaluate the normalization constant ${\cal N}$ and the average
energy $\bar E$.  Figure \ref{fig:Fig1.eps} plots the average final energy 
$\bar E$ for an $\alpha$ particle in an equimolar DT plasma with an 
electron density $n_e = 1.0 \times 10^{25} \, {\rm cm}^{-3}$. The figure 
displays $\bar E$ as a function of the ion temperature $T_\smI$ for 
various electron temperatures $T_e$. 

\section{Formal Solution}
\label{sec:evolution}

\subsection{A Homogeneous and Isotropic Source}

We shall assume that the background plasma parameters, such as its
density and temperatures, change very little over distances that are 
large in comparison with the stopping distance of the charged impurity
particles, and that the plasma parameters also change very little during
the stopping time. Thus the plasma is treated as homogeneous and static.
In addition, we assume that the sources of the impurity particles are
distributed uniformly in space and that they emit the impurity particles
isotropically with a definite energy $E_0$.  For example, the fusion 
process in a homogeneous DT plasma produces $\alpha$ particles uniformly
in space and isotropically in angle with an initial energy of 
$E_0 = 3.54$ MeV. Thus, instead of considering the motion of a single
test particle, we compute  energy partitions and final states of 
charged particles emitted isotropically with a definite energy $E_0$ from
a uniform distribution of sources.  This greatly simplifies
the problem in that we can employ the homogeneous Fokker-Planck 
Eq.~(\ref{fphomoE}) except that it is now modified to include a time-varying
source of particles of energy $E_0$:
\begin{equation}
  \left\{\frac{\partial}{\partial t} 
  - 
  \frac{2}{m v}\, \frac{\partial}{\partial E}\,
  E \left[ {\cal A}(E) + \langle T {\cal A}(E)\rangle\,
  \frac{\partial}{\partial E}
  \right] \right\} \, f(E,t) 
  = 
  \delta\left( E - E_0 \right) \, s(t) \,.
\label{inhomo} 
\end{equation}
The number and energy densities, $n(t)$ and ${\cal E}(t)$,  
are simply given by removing the spatial volume integrations from the
previous definitions (\ref{number}) and (\ref{energy}).
The inhomogeneous Fokker-Planck equation (\ref{inhomo}) gives the time
variations of these quantities:
\begin{equation}
  \dot n(t) 
  =
 \int \frac{d^3p}{(2\pi\hbar)} \,
 \delta\left( E - E_0 \right) \, s(t) 
  =
  \frac{s(t)}{2 \pi^2 \hbar^3} \, \sqrt{2 m^3 E_0} \,,
\label{ndot}
\end{equation}
and
\begin{eqnarray}
  \dot {\cal E}(t) 
  &=& 
  E_0 \, \dot n(t) - \int\frac{d^3{\bf p}}{(2\pi\hbar)^3}\, 
   v \,  \left\{ \left[{\cal A}_\smI(E) + {\cal A}_e(E) \right] +
\left[T_\smI \, {\cal A}_\smI(E) + T_e \,  {\cal A}_e(E) \right] 
  \frac{\partial}{\partial E} \right\}  \, f(E,t) \,.
\nonumber\\
&&
\label{Edot}
\end{eqnarray}
When the impurity source $s(t)$ is turned on and then attains  a constant fixed
value $s_0$, the number density $n(t)$ eventually increases linearly in time,
\begin{eqnarray}
  n(t) 
  &=& 
  \int_{-\infty}^t dt' \, \dot n(t') 
  =  
  \frac{\sqrt{2 m^3 E_0} }{2 \pi^2 \hbar^3}\, 
  \int_{-\infty}^t dt' \, s(t') 
\nonumber\\[5pt]
     &=& \dot n_\infty \, t  + {\rm constant} \,,
\label{numberr}
\end{eqnarray}
where
\begin{equation}
\dot n_\infty = \frac{s_0 }{2 \pi^2 \hbar^3} \, \sqrt{2 m^3 E_0} \,.
\label{ndott}
\end{equation}

\subsection{Asymptotic Solution to the Inhomogeneous Problem}
\label{problem}

We turn now to obtain the asymptotic solution to
(\ref{inhomo}) satisfying the initial condition that there are no
impurity particles in the distant past.

As a first step in obtaining the asymptotic solution of the
inhomogeneous Fokker-Planck equation (\ref{inhomo}), we set
\begin{equation}
f(E,t) = \exp\left\{- S(E) /2 \right\} \, g(E,t) \,.
\end{equation}
Multiplying the resulting Fokker-Planck equation by $\exp\{S(E)/2\}$
on the left yields a similarity transformation that converts the
(velocity $\sim$ momentum) differential operator structure in
Eq.~(\ref{inhomo}) into
\begin{equation}
H = - \left[ \frac{\partial}{\partial {\bf p}} \cdot \hat {\bf v}
         - \frac{v{\cal A}(E)}{2  \langle T{\cal A}(E) \rangle} \right] 
\, \frac{ \langle T{\cal A}(E) \rangle}{ v } \,  
\left[\hat{\bf v} \cdot \frac{\partial}{\partial{\bf p}}
    + \frac{v {\cal A}(E)}{2  \langle T{\cal A}(E) \rangle} \right] \,,
\label{Ham}
\end{equation}
so that the new Fokker-Planck equation now appears as
\begin{equation}
\left\{ \frac{\partial}{\partial t} + H \right\} g(E,t) =
   \delta\left( E - E_0 \right) e^{S(E_0)/2} \, s(t) \,.
\label{Hameq}
\end{equation}
Incorporating the boundary condition that the solution vanishes
initially, the inhomogeneous differential equation (\ref{Hameq}) has a
formal solution:
\begin{equation}
     g(E,t) = \int_{-\infty}^t dt' \, e^{-H(t-t')} \, 
              \delta\left( E - E_0 \right) \, e^{S(E_0)/2} \, s(t') \,.
\label{fromal}
\end{equation}

Because of the operator nature of the formal solution
(\ref{fromal}), it is convenient to view functions in momentum
space as vectors in an abstract real vector space and 
define an inner product by
\begin{equation}
  (\psi,\chi) = \int \frac{d^3{\bf p}}{(2\pi\hbar)^3} \, 
  \psi({\bf p}) \, \chi({\bf p}) \,.
\end{equation}
With obvious partial integrations, it is straightforward to verify
that $H$ considered as an operator on this function space is Hermitian
with this definition of the inner product. 

In view of our previous work, it is easy to check that 
\begin{equation}
  \phi({\bf p}) 
  = 
  {\cal N}^{1/2} \, \exp\left\{ - S(E) / 2 \right\}
\label{phiformal}
\end{equation}
now appears as a zero mode of the operator $H$,
\begin{equation}
  H \, \phi = 0 
\end{equation}
that has unit normalization, 
\begin{equation}
(\phi,\phi) = 1 \,.
\end{equation}
Except for this zero mode function, the remaining spectrum of $H$ is
positive.  This is true because, for any function $\psi({\bf p})$,
\begin{equation}
(\psi, H \psi) = \int \frac{d^3{\bf p}}{(2\pi\hbar)^3} \, 
   \frac{ \langle T {\cal A}(E) \rangle}{v} \left\{
\left[ \hat {\bf v} \cdot \frac{\partial}{\partial {\bf p}} +
\frac{v \, {\cal A}(E)}{2\langle T {\cal A}(E) \rangle} \right]
      \psi({\bf p}) \right\}^2 \geq 0 \,,
\label{positive}
\end{equation}
since an examination of our results for the ${\cal A}$ coefficients
shows that $ \langle T {\cal A}(E) \rangle \geq 0$. The equality in 
Eq.~(\ref{positive}) holds only if
\begin{equation}
\left[ \hat {\bf v} \cdot \frac{\partial}{\partial {\bf p}} +
\frac{v \, {\cal A}(E)}{2\langle{T\cal A}(E)\rangle} \right]
      \psi({\bf p}) = 0 \,.
\end{equation}
The spherically symmetric solution 
$\psi({\bf p}) = \psi(|{\bf p}|) $ is clearly the previous
zero mode function $\psi({\bf p}) = \phi(E)$. Hence within the
class of isotropic solutions --- the only class that is relevant
to our work --- there are no other zero modes of $H$ and all its 
other eigenvalues are positive. Since the operator $H$ is Hermitian,
\begin{equation}
\phi \, H = 0 \,.
\end{equation}
In view of this adjoint equation, it follows that
\begin{equation}
\phi \, e^{-H (t-t') } = 1 \,,
\end{equation}
as one easily verifies by taking the time derivative.

Except for this zero mode, we have shown that the other eigenvalues
of the Hermitian operator $H$ are positive.  This positivity
constraint must be obeyed, for otherwise the Fokker-Planck would have
diverging ``runaway'' solutions at large times.  The operator that
projects out the zero mode is obviously the outer product of the zero
mode vector with itself,
\begin{equation}
  P = \phi \otimes \phi \,,
\label{P}
\end{equation}
and we write the complement operator as
\begin{equation}
  Q = 1 - P \ ,
\label{Q}
\end{equation}
where the first term in (\ref{Q}) is the unit operator on the function
space. By definition, the operator $P$ acts on an arbitrary
function $\psi$ as
\begin{equation}
  P\,\psi({\bf p}) 
  = 
  \Big(\phi \otimes \phi\Big)\psi({\bf p}) 
  =
  \phi({\bf p})\,(\phi,\psi) \ .
\label{Pop}
\end{equation}
We now see that  the unit operator in the form $P + Q$ acting on $g(E,t)$ in
(\ref{fromal}) produces
\begin{eqnarray}
  g(E,t) 
  &=& 
    \phi({\bf p}) \, \int \frac{d^3p'}{(2\pi\hbar)^3}\, 
  \phi({\bf p}') \delta\left( E' - E_0 \right) \, 
  e^{S(E_0)/2} \,  \int_{-\infty}^t dt' \, s(t') 
\nonumber\\
&& \qquad + 
\int_{-\infty}^t dt' \, e^{-H(t-t')} \, Q
         \delta\left( E - E_0 \right) \, e^{S(E_0)/2} \, s(t') \,.
\label{result}
\end{eqnarray}
The momentum integral in the first term of (\ref{result}) is easy to
evaluate, 
\begin{equation}
  \phi({\bf p}) \, \int \frac{d^3{\bf p}'}{(2\pi\hbar)^3} \, 
  \phi({\bf p}') \delta\left( E' - E_0 \right) \, e^{S(E_0)/2} =
{\cal N} \, e^{-S(E)/2} \, \frac{\sqrt{2 m^3 E_0}}{2 \pi^2 \hbar^3} \,.
\label{noted}
\end{equation}
As for the second term, since the operator $Q$ selects out
the positive eigenvalues of $H$, an integration by parts can be
performed to produce
\begin{eqnarray}
  \int_{-\infty}^t dt' \, e^{-H(t-t')} \, Q
  \delta\left( E - E_0 \right) \, e^{S(E_0)/2} \, s(t') 
  &=& 
  \frac{1}{H} \, Q \delta\left( E - E_0 \right) \, e^{S(E_0)/2} \, s(t) 
\nonumber\\
  - && \!\!\!\!\!\!\!\!\! \int_{-\infty}^t dt' \, e^{-H(t-t')} \,
  \frac{1}{H} \, Q \delta\left( E - E_0 \right) \, e^{S(E_0)/2} \, 
  \dot s(t') \,.
\nonumber\\
\label{Pong}
\end{eqnarray}
We now assume that the source $s(t)$ is adiabatically turned on
and attains the constant value $s(t) = s_0$ at late times. In the
asymptotic limit, the rate $\dot s(t)$ is therefore vanishingly small
and the second term in Eq.~(\ref{Pong}) may be neglected. We may also
replace $s(t)$ by its asymptotic value $s_0$ in the first line.
Hence, upon multiplying $g(E,t)$ by $\exp\{- S(E)/2 \} $ to return to the
function $f(E,t)$, we obtain
\begin{equation}        
  f(E,t) = {\cal N} \, e^{-S(E)} \, \frac{\sqrt{2 m^3 E_0}}{2 \pi^2 \hbar^3} \,
  \int_{-\infty}^t dt' \, s(t') + \bar f(E) \,,
\label{fEt}
\end{equation}
where 
\begin{equation}
\bar f(E) = e^{-S(E)/2} \, 
\frac{1}{H} \, Q \delta\left( E - E_0 \right) \, e^{S(E_0)/2} \, s_0 \,.
\label{barf}
\end{equation}
Upon integrating (\ref{ndot}), we can write this asymptotic late time
solution more suggestively as
\begin{equation}
  f(E,t) = n(t) f_\infty(E)  + \bar f(E) \,,
\label{asymp}
\end{equation}
with $f_\infty(E)={\cal N} \, e^{-S(E)}$.  We emphasize that
expression (\ref{asymp}) is the asymptotic late-time solution to the
inhomogeneous Fokker-Planck equation since the term involving
the derivative $\dot s(t)$ is omitted.

\subsection{Energy Deposition}

\begin{figure}[t]
\includegraphics[scale=0.5]{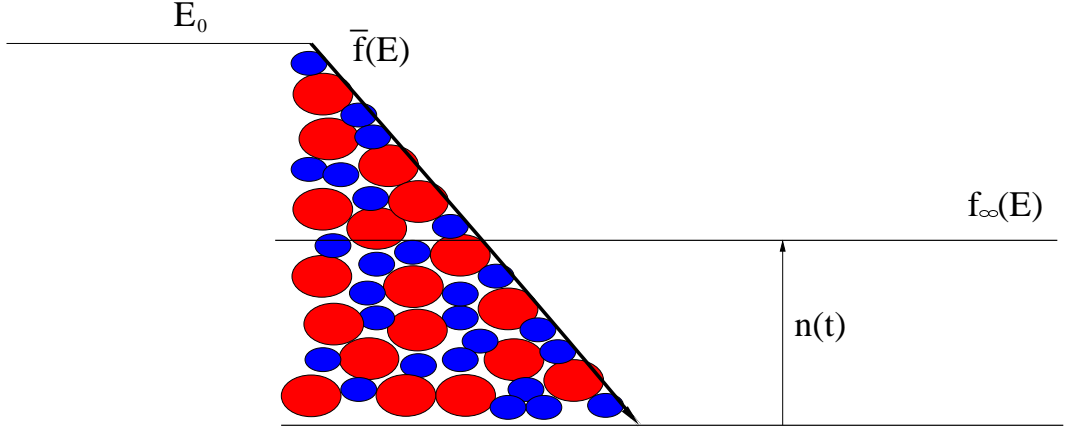}
\caption{\captionskip  
  The waterfall analogy: the small blue rocks 
  represent the plasma electrons while the larger red rocks are
  the plasma ions. The motion of the `water' represents the
  evolution of the impurity ions that are injected into the
  background plasma. As `water' falls down the electron-ion slope
  at a constant rate determined by $\bar f(E)$, energy is deposited
  into electrons and ions. At the bottom of the fall is a 
  lake into which the excess `water' drains and whose height $n(t)$
  rises linearly with time.
}
\label{fig:name}
\end{figure}

Before presenting an explicit version of the formal solution
(\ref{asymp}), we pause to describe its physical interpretation
and its relation to the ways in which the stopping charged
particle deposits its energy to the background plasma.  
At  large times, the phase-space density has the time-independent 
contribution $f_\infty(E) = {\cal N} \,\exp\{-S(E)\}$ into which 
any set of initial test particles must relax, and the first term 
of (\ref{asymp}) describes this distribution normalized to the 
correct density $n(t)$.  There
remains a time-independent part $\bar f(E)$ that describes the stationary
process of particles losing energy to the background electrons and ions
as particles pass through ``energy bins'' from the initial energy
$E_0$ to the final asymptotic distribution.  The situation described
here can be pictured as the flow of water over a rocky waterfall that
slows the motion of the water as it descends.  The initial rate of
flow of the river corresponds to the rate $\dot n(t)$; the height $h$
of the waterfall giving a potential energy proportional to $g h$
corresponds to the initial energy $E_0$. The energy dissipated in the
fall corresponds to the energies lost to the ions and electrons.  The
final flow into a horizontal lake corresponds to the build up of the
particles in their final distribution described by $f_\infty(E)$.
This analogy is depicted in Fig.~\ref{fig:name}. 

\subsubsection{Energy Splitting}

Upon inserting the form (\ref{asymp}) into Eq.~(\ref{Edot}), we
can identify the asymptotic constant rates of energy loss as 
\begin{equation}
\bar E \, \dot n_\infty = 
   \left[ E_0 - E_\smI - E_e \right] \, \dot n_\infty \,,
\label{obvious}
\end{equation}
in which
\begin{eqnarray}
\frac{E_\smI}{E_0} &=& \frac{1}{\dot n_\infty \, E_0} 
\int\frac{d^3{\bf p}}{(2\pi\hbar)^3} \, v \,   {\cal A}_\smI(v) 
 \, \left[ 1 + T_\smI\,
   \frac{\partial}{\partial E} \right]  \, \bar f(E) \,,
\label{Ifrac}
\end{eqnarray}
and
\begin{eqnarray}
\frac{E_e}{E_0} &=& \frac{1}{\dot n_\infty \, E_0} 
\int\frac{d^3{\bf p}}{(2\pi\hbar)^3} \, v \  {\cal A}_e(v) 
\, \left[ 1 + T_e\,
   \frac{\partial}{\partial E} \right]  \, \bar f(E) \,.
\label{efrac}
\end{eqnarray}
Thus Eqs.~(\ref{Ifrac}) and (\ref{efrac}) are the constant
fractions of the original energy $E_0$ deposited into ionic
energy $E_\smI$ and electronic energy $E_e$ --- the energy losses
analogous to those of the water passing through the rocky
waterfall.  

\subsubsection{Plasma Heating and Energy Exchange}
\label{exchange}{exchange}

The original energy goes into energies lost to the ions and
electrons, with the remainder the average final energy $\bar E$
of an impurity particle. For a background plasma with the ions and
electrons at a common temperature $T$, $\bar E = 3 T / 2$, and
the result (\ref{obvious}) becomes obvious. 

When the electrons and ions have the same temperature $T = T_e =
T_\smI$, the slowing down of fast particles in the plasma gives a
steady-state heating rate per unit volume 
${\cal P} = [ E_\smI + E_e ] \, \dot n_\infty$. 
This heating raises the temperature $T$ of the plasma, but in
most cases, the rate of this heating is small in comparison with
the slowing down time of the fast impurity particles, and so our
quasi-steady-state computation is valid, with the temperature
treated as a slowing varying function in our formulae.

When the electrons and ions have different temperatures $T_e$ and
$T_\smI$, the situation may be quite different.  In addition to
the overall plasma heating ${\cal P}$, the final ensemble of the 
impurity particles works to bring the electrons and ions to a common
temperature. Returning to Eq.~(\ref{Edot}), we see that the final
ensemble contribution produces energy density transfer rates to
the ions and electrons given by
\begin{eqnarray}
\dot {\cal E}_\smI(t) &=& + \int\frac{d^3{\bf p}}{(2\pi\hbar)^3}
\,  v \,  {\cal A}_\smI(v) \left[ 1 + T_\smI \, 
\frac{\partial}{\partial E}  \right]  \, {\cal N} \, 
\exp\left\{ - S(E) \right\} \, n(t) \,,
\label{idot}
\end{eqnarray}
and
\begin{eqnarray}
\dot {\cal E}_e(t) &=& + \int\frac{d^3{\bf p}}{(2\pi\hbar)^3}
\,  v \,  {\cal A}_e(v) \left[ 1 + T_e \, 
\frac{\partial}{\partial E}  \right]  \, {\cal N} \, 
\exp\left\{ - S(E) \right\} \, n(t) \,.
\label{edot}
\end{eqnarray}
Carrying out the energy derivatives yields
\begin{eqnarray}
\dot {\cal E}_\smI(t) &=& - \left( T_\smI - T_e \right) \,
C^\alpha_{\smI \, e} \,,
\end{eqnarray}
and
\begin{eqnarray}
\dot {\cal E}_e(t) &=& - \left( T_e - T_\smI \right) \,
C^\alpha_{e \, \smI} \,,
\end{eqnarray}
where
\begin{eqnarray}
C^\alpha_{\smI \, e} = C^\alpha_{e \, \smI} 
 &=& n(t) \,  \int\frac{d^3{\bf p}}{(2\pi\hbar)^3}
\,  v \,  \frac{{\cal A}_\smI(v) \, {\cal A}_e(v)}{\langle T
 {\cal A}\rangle } \,  {\cal N} \, \exp\left\{ - S(E) \right\} \,.
\end{eqnarray}
Since
\begin{equation}
\dot {\cal E}_\smI(t) + \dot {\cal E}_e(t) = 0 \,,
\end{equation}
there is no net heating of the plasma.  This process only brings
the ions and electrons to a common temperature. 

In the absence of the impurity particle quasi-static equilibrium ensemble,
the thermal relaxation rate coefficients are well approximated\footnote{
This is the sum of Eqs.~(12.44) and (12.57) in BPS \cite{bps} as quoted in
Eq.~(12.12) except that a simple transcription error was made in the sum
quoted in BPS in that the $- \gamma - 2$ in Eq.~(12.12) should be replaced
by $-\gamma - 1$.}
by
\begin{equation}
C_{\smI \, e} = C_{e \, \smI} = \frac{\kappa_e^2}{2\pi} \, \omega_\smI^2
    \sqrt{\frac{m_e}{2\pi \, T_e}} \, \frac{1}{2} \left\{
     \ln\left( \frac{8 T_e^2}{\hbar^2 \, \omega_e^2} \right)
         - \gamma - 1 \right\} \,.
\end{equation}
Here
\begin{equation}
\kappa_e^2 = \frac{e^2 \, n_e}{T_e} 
\end{equation}
is the squared electron Debye wave number, and
\begin{equation}
\omega_a^2 = \frac{e_a^2 n_a}{m_a} 
\end{equation}
is the definition of the squared plasma frequency for particle $a$,
with the electron squared plasma frequency $\omega_e^2$ specified by 
$a=e$, while the total squared ionic plasma frequency $\omega_\smI^2$ 
is the sum over all the ions in the plasma
\begin{equation}
\omega_\smI^2 = {\sum}_i \, \omega_i^2 \,.
\end{equation}
In numerical terms, for an equimolar DT plasma,
\begin{equation}
C_{\smI \,e} = 3.13 \times 10^{-26} \, n_e^2 \, T_e^{-3/2} \,
    \left\{ \ln\left( 5.80 \times 10^{27} \, \frac{T_e^2}{n_e} \right)\   
        - 1.58 \right\} \, {\rm cm}^{-3} \, {\rm ps}^{-1} \,,
\end{equation}
in which the electron density $n_e$ is measured in ${\rm cm}^{-3}$,
the electron temperature $T_e$ in keV, and the overall units are
$(1.0 \, {\rm cm}^{-3}) / ( 1.0 \times 10^{-12} \, {\rm sec}) $ as indicated.

\begin{figure}[t!]
\includegraphics[scale=0.45]{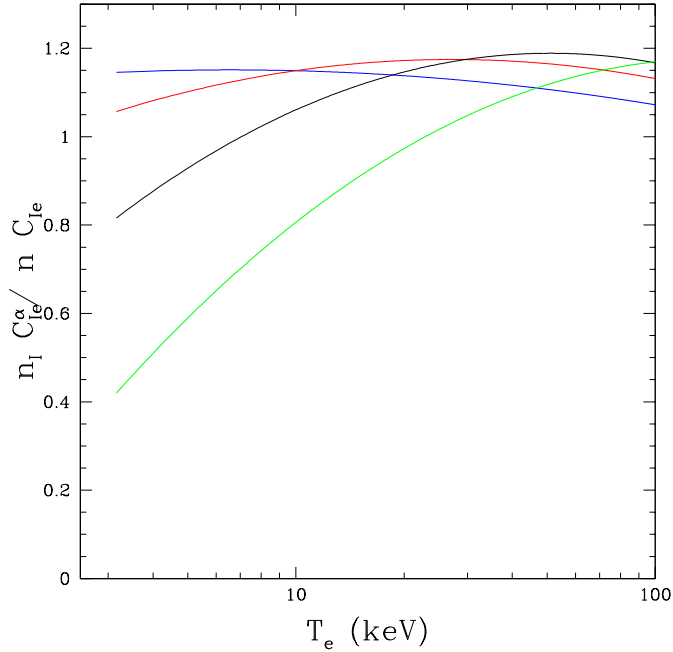}
\vskip-0.4cm 
\caption{\footnoteskip  
The ratio $n_\smI \, C^\alpha_{\smI \, e} / n \, C_{\smI \, e} $ as a function
of the electron temperature for an equimolar DT plasma 
with an electron density of $1.0 \times 10^{24} \, {\rm cm}^{-3}$ for 
ion temperatures of 3 keV (blue), 10 keV (red), 30 keV (black), and
100 keV (green). 
}
\label{fig:file}
\end{figure}

The total rate coefficient for electron-ion thermal relaxation is the sum
$C_{\smI \, e} + C^\alpha_{\smI \, e}$. It is of interest to compare
$C^\alpha_{\smI \, e}$ to $C_{\smI \, e}$. Since $C^\alpha_{\smI \, e}$
is proportional to the number of impurity particles that come into their
final equilibrium state $n(t)$, this comparison can be made independent of
this density by evaluating the ratio of $C^\alpha_{\smI \, e} / n(t)$
to $C_{\smI \, e} / n_\smI$.  In Fig.~\ref{fig:file} we plot this dimensionless
ratio as a function of the electron temperature $T_e$ for various values
of the ion temperature $T_\smI$ ranging from 3 keV to 100 keV at an
electron density $n_e = 1.0 \times 10^{24} \, {\rm cm}^{-3}$. Explicit
calculation shows that the dependence of this ratio upon the electron
density $n_e$ is weak.  As $n_e$ is increased from 
$1.0 \times 10^{24} \, {\rm cm}^{-3}$ to 
$1.0 \times 10^{26} \, {\rm cm}^{-3}$, the greatest change in the ratio
occurs for $T_\smI \gg T_e$\,: for $T_\smI = 100$ keV and $T_e = 3$ keV,
the ratio increases by 20\%.

We must add the caveat, already noted in the Introduction, that the
discussion that we have just made applies only to the case in which
the final alpha particle population is not large. Hence, although in
some cases the ratios shown in Fig.~\ref{fig:file} are of order one,
the net effect of this new mechanism must be relatively small.

\subsection{Results in Terms of $dE/dx$}

The results (\ref{Ifrac}), (\ref{efrac}), (\ref{idot}), and (\ref{edot})
all have the generic structure
\begin{eqnarray}
\tilde E_{\smI,e} &=&
\int\frac{d^3{\bf p}}{(2\pi\hbar)^3} \, v \  {\cal A}_{\smI,e}(v) 
\, \left[ 1 + T_{\smI,e} \,
   \frac{\partial}{\partial E} \right]  \, \tilde f(E) \,.
\label{frac}
\end{eqnarray}
Here we may write
\begin{equation}
 \frac{\partial}{\partial E} \, \tilde f(E) = 
\frac{1}{2E} \, {\bf p} \cdot \frac{\partial}{\partial {\bf p}} \tilde f(E) \,,
\end{equation}
use ${\bf p} = m {\bf v}$, and integrate by parts to obtain
\begin{eqnarray}
\tilde E_{\smI,e} &=&
\int\frac{d^3{\bf p}}{(2\pi\hbar)^3} \,  \tilde f(E) 
\, \left[ v - \frac{T_{\smI,e}}{m} \,
   \frac{\partial}{\partial {\bf v}} \cdot \hat{\bf v} \right] 
{\cal A}_{\smI,e}(v) \,.
\label{ffrac}
\end{eqnarray}
As remarked in the Introduction [Eq.~(\ref{stopping})], the expression
to the right of $\tilde f(E)$ in the integrand above is just
$ v \, dE_{\smI,e}(E) /dx $. The integrand does not depend upon the direction
of ${\bf p}$.  Thus the angular integration simply provides a factor of
$4\pi$. Using $v dp = dE$, we now have
\begin{eqnarray}
\tilde E_{\smI,e} &=&
 \int_0^\infty dE \,\frac{dE_{\smI,e}(E)}{dx} \, \frac{p^2}{2 \pi^2 \hbar^3}
  \, \tilde f(E) \,.
\label{ffracc}
\end{eqnarray}
Thus all of our results involve a factor of the stopping power
$d E_{\smI,e}/dx$ for the ions or for the electrons, but the integration 
weight involves a more subtle function than those in the naive formulae
(\ref{eion}) and (\ref{eelectron}) given in the Introduction.

\section{Explicit Solution}
\label{explicit}

\subsection{General Development}
\label{gendev}

We turn now to the explicit 
construction of the function $\bar f(E)$ from the formal expression
(\ref{barf}). We start by multiplying Eq.~(\ref{barf}) by the (velocity
$\sim$ momentum) differential operator structure in
Eq.~(\ref{inhomo}). Passing this operator through the factor $\exp\{ -
S(E)/2\}$ is equivalent to the similarity transformation that converts
it into the operator $H$.  Hence,
\begin{equation}
  -\frac{\partial}{\partial {\bf p}} \cdot \hat {\bf v} \, \left[
  {\cal A} +  \langle T{\cal A}(E) \rangle \, \frac{\hat {\bf v}}{v}
  \cdot \frac{\partial}{\partial{\bf p}} \right]  \, \bar f(E) 
  = 
  e^{-S(E)/2}\,Q\,e^{+S(E)/2} \,\delta\left( E - E_0 \right) \, s_0 \,,
\label{fbarr}
\end{equation}
and remembering Eqs.~(\ref{usefulEQ}), we see that this is
equivalent to
\begin{eqnarray}
  -\frac{\partial}{\partial v}  \,\frac{v^2}{m} \left[ {\cal A} +  
  \langle T{\cal A}(E) \rangle \, 
  \frac{\partial}{\partial E} \right]  \, \bar f(E) 
  &=& 
  v^2 e^{-S(E)/2} \, Q \, e^{+S(E)/2}\,\delta\left( E - E_0 \right) \, s_0
\nonumber\\
  &=& 
  \delta\left( E - E_0 \right)\,\frac{2E_0}{m} \, s_0 - 
  v^2 {\cal N} \,  e^{- S(E)}\,\frac{\sqrt{2 m^3 E_0}}{2\pi^2
  \hbar^3}\, s_0 \,.
\nonumber\\
\label{ffbarr}
\end{eqnarray}
In the second equality we employed the definitions (\ref{P}) and
(\ref{Q}) of the operators $P$ and $Q$ and in the last line used the
result (\ref{noted}). Obviously, a trivial first integral of this
differential equation exists.  Since the constant of integration must
be chosen to make $\bar f(E)$ vanish at large $E$, this first integral
reads
\begin{eqnarray}
\left[ {\cal A}(E) +  \langle T{\cal A}(E) \rangle
 \frac{\partial}{\partial E} \right] \bar f(E)  
&=&   \frac{s_0}{ E }
 \sqrt{\frac{mE_0}{2 }} \left\{  \theta\left( E_0 - E \right)  
 -    \int_E^\infty dE' \, \frac{m \, \sqrt{2mE'} }{2 \pi^2 \hbar^3} \, 
  {\cal N}  e^{-S(E')}  \right\} ,
\nonumber\\
&&
\label{ffbarrr}
\end{eqnarray}
where $\theta(x)$ is the unit step function that vanishes for $x<0$.
Note that, in view of the normalization (\ref{onenorm}), 
\begin{equation}
 \int_0^\infty dE' \frac{m \, \sqrt{2mE'} }{2 \pi^2 \hbar^3} \, 
  {\cal N} \, e^{-S(E')} = 
   \int \frac{d^3{\bf p}'}{(2\pi\hbar)^3} \,  {\cal N} \, e^{-S(E')} 
    = 1 \,,
\label{sum}
\end{equation}
and so the sum of the terms in the curly braces in Eq.~(\ref{ffbarrr})  
vanishes when $E \to 0$.  This is in accord with the 
fact that these terms on the right of Eq.~(\ref{ffbarrr}) were 
produced by the integral of a derivative on the left-hand side of 
Eq.~(\ref{ffbarr}), a derivative of a quantity
that vanishes at both $E=0$ and $E= \infty$.  Moreover, since
the curly braces vanishes at $E=0$, the right-hand side of
Eq.~(\ref{ffbarrr}) is finite at this end point as it must be.

At this juncture, it is convenient to remember the definition 
(\ref{ndott}) of $\dot n_\infty$, which can be expressed as
\begin{equation}
\sqrt{\frac{m E_0}{2}} \, s_0  =
\frac{\pi^2 \hbar^3}{ m  E_0} \, E_0 \, \dot n_\infty \,,
\end{equation}
and to simplify the notation by writing 
\begin{equation}
\overline{\cal N} =  \frac{m \, \sqrt{2m} }{2 \pi^2 \hbar^3} \, {\cal N} \,,
\end{equation}
so that we have 
\begin{equation}
 \int_0^\infty dE'  \sqrt{E'} \,\, 
  \overline{\cal N} \,\, e^{-S(E')}  = 1 \,.
\label{summ}
\end{equation}
Thus, Eq.(\ref{ffbarrr}) now reads:
\begin{eqnarray}
\left[ {\cal A}(E) +  \langle T{\cal A}(E) \rangle
 \frac{\partial}{\partial E} \right]
\, \frac{\bar f(E)}{E_0 \, \dot n_\infty}  
&=&   \frac{\pi^2 \hbar^3 }{ m E_0 E } \, 
 \left\{  \theta\left( E_0 - E \right) \, 
 -    \int_E^\infty dE'  \sqrt{E'} \,
  \overline{\cal N} \, e^{-S(E')}  \right\} \,.
\nonumber\\
&&
\label{fffbarrr}
\end{eqnarray}

To solve this differential equation, we set
\begin{equation}
\bar f(E) = e^{ - S(E) } \, \bar g(E) \,,
\label{gbar}
\end{equation}
because then
\begin{equation}
\left[ {\cal A}(E) +  \langle T{\cal A}(E) \rangle
 \frac{\partial}{\partial E} \right] \, \bar f(E) =
e^{ - S(E) } \,  \langle T{\cal A}(E) \rangle \,
 \frac{\partial}{\partial E} \, \bar g(E) \,.
\label{solve}
\end{equation}
Since the integrating factor involves  $\exp\{ + S(E) \}$,
which exponentially increases without bound as the energy increases,
to obtain a finite well-defined result we must integrate over the
range $E^\prime=0$ to $E^\prime=E$ and obtain
\begin{eqnarray}
  \frac{\bar g(E)}{E_0 \, \dot n_\infty}  &=& 
  \frac{\pi^2 \hbar^3 }{ m E_0 } \, \int_0^E \frac{dE'}{E'} \,
  \frac{e^{ + S(E') }}{ \langle T{\cal A}(E') \rangle } \,
  \left\{  \theta\left( E_0 - E' \right) \, 
  -\int_{E'}^\infty dE''\sqrt{E''} \,\, \overline{{\cal N}} \, \, e^{-S(E'')}  
  \right\} .
\label{barg}
\end{eqnarray}

\subsection{Energy Fractions and $dE/dx$}
\label{EfracdE/dx}

The customary expressions for the energy fractions in terms of
the stopping power [Eqs.~(\ref{eion}) and (\ref{eelectron})] emerge
for low temperatures.  To see this, we note that in this case, the
energy integration range is very large on the scale of the temperature,
and that the work of Appendix \ref{sec:limits} shows that the electron
contribution dominates over most of this range so that we may approximate
\begin{equation}
S(E) \simeq \frac{E}{T_e} \,.
\end{equation}
Moreover, the sum rule (\ref{summ}) implies that the terms in the curly
braces in Eq.~(\ref{barg}) cancel when $E' \simeq \tilde E $, where 
$\tilde E$ is a lower energy limit that is on the order of the
electron temperature $T_e$. On the other hand, the integral in the 
curly braces in Eq.~(\ref{barg}) is exponentially small when the integration
variable $E'$ is somewhat larger that the electron temperature $T_e$.  
Hence, in the low temperature case, Eqs.~(\ref{gbar}) and (\ref{barg}) 
provide the approximate solution
\begin{eqnarray}
\bar f(E) &\simeq& 
\dot n_\infty \,  \frac{\pi^2 \hbar^3 }{ m } \, \int_0^E \frac{dE'}{E'} \,
\exp\left\{ - \frac{1}{T_e} \, \left( E - E' \right) \right\} \,
\frac{1}{ \langle T{\cal A}(E') \rangle } \,
 \theta\Big( E_0 - E' \Big) \,  \theta\Big( E' - \tilde E \Big) \,. 
\nonumber\\
&&
\label{bbarg}
\end{eqnarray}
Repeatedly using
\begin{equation}
\exp\left\{ - \frac{1}{T_e} \, \left( E - E' \right) \right\} =
T_e \, \frac{d}{dE'} \,
\exp\left\{ - \frac{1}{T_e} \, \left( E - E' \right) \right\} \,,
\end{equation}
and repeatedly integrating by parts, shows that the leading term
in the low temperature case is given by the upper-limit contribution
of the first term in this sequence:
\begin{eqnarray}
f(E) &\simeq& 
\dot n_\infty \,  \frac{\pi^2 \hbar^3 }{ m } \, 
\frac{T_e}{E} \, \frac{1}{ \langle T{\cal A}(E) \rangle } \,
 \theta\Big( E_0 - E \Big) \,  \theta\Big( E - \tilde E \Big) \,. 
\label{bbargg}
\end{eqnarray}

Placing this approximate result in the generic form (\ref{ffracc})
to evaluate the energy fractions (\ref{Ifrac}) and ({\ref{efrac})
yields
\begin{equation}
\frac{E_\smI}{E_0} \simeq \int_{\tilde E}^{E_0} \frac{dE}{E_0} \, 
\frac{T_e}{ \langle T{\cal A}(E) \rangle } \,
  \frac{dE_\smI}{dx}(E) \,,
\label{AIfrac}
\end{equation}
and
\begin{equation}
\frac{E_e}{E_0} \simeq \int_{\tilde E}^{E_0} \frac{dE}{E_0} \, 
\frac{T_e}{ \langle T{\cal A}(E) \rangle } \,
  \frac{dE_e}{dx}(E) \,.
\label{Aefrac}
\end{equation}
In the low temperature case, the generic relation (\ref{stopping})
gives $ dE / dx \simeq {\cal A}$ and so
\begin{equation}
\frac{T_e}{ \langle T{\cal A}(E) \rangle } \simeq
\frac{T_e}{T_\smI \, dE_\smI/dx(E) + T_e \, dE_e/dx(E) } \,.
\end{equation}
In the equal temperature case,
\begin{equation}
\frac{T_e}{ \langle T{\cal A}(E) \rangle } \to
\frac{1}{ dE/dx(E) } \,,
\end{equation}
and, setting $E_0 \to 0$,  the low temperature expressions 
(\ref{AIfrac}) and (\ref{Aefrac}) reduce to the commonly used 
Eqs.~(\ref{eion}) and (\ref{eelectron}) discussed in the Introduction.

\subsection{Equal Electron and Ion Temperatures}
\label{equaltemps}

The case in which the ions and electrons have the same temperature,
$T_\smI=T_e=T$, is simple in several respects. First of all, it
is physically simpler because the final distribution of the
stopping charged particles is the Maxwell-Boltzmann thermal
equilibrium distribution of the background plasma, 
\begin{equation}
\exp\left\{-S(E)\right\} = \exp\left\{-\frac{E}{T} \right\} \,.
\label{eeasy}
\end{equation}
Thus, the energy transfer processes (\ref{idot}) and (\ref{edot})
do not appear because, with Eq.~(\ref{eeasy}) holding,  the
combinations in the square brackets in these equations annihilate  
$\exp\{-S(E)\}$. Thus, only the energy partitions $E_\smI$ and
$E_e$ need to be examined, and these obey the obvious sum rule 
\begin{equation}
\frac{3}{2} \, T = E_0 - E_\smI - E_e \,,
\end{equation}
to which Eq.~(\ref{obvious}) reduces.  
Secondly, it is mathematically simpler because there is no need
to find an explicit solution to Eq.~(\ref{barg}) because 
Eq.~(\ref{fffbarrr}) reduces to 
\begin{eqnarray}
  \left[ 1 + T \, \frac{\partial}{\partial E} \right]\, 
  \frac{\bar f(E)}{E_0 \dot n_\infty}  
  &=&  
  \theta\left( E_0 - E \right) \, \frac{1}{ E {\cal A}(E)} \,  
 \frac{\pi^2 \hbar^3}{m E_0}  
\nonumber\\[5pt]
  && 
  -\frac{1}{ E_0 E \, {\cal A}(E)} \, 
\left( \frac{2\pi \hbar^2}{mT} \right)^{3/2} \, 
  \int_E^\infty dE'  \, \sqrt{2mE'} \,  e^{-\beta E'} \,.
\label{ffbarrrOneT}
\end{eqnarray}
The operation in the square brackets that acts on $\bar f(E)$ on
the left-hand side of this equation is just that which appears
in the energy partitions (\ref{Ifrac}) and (\ref{efrac}).

Placing this expression into the energy partitions
Eqs.~(\ref{Ifrac}) and (\ref{efrac}) and changing the momentum
integration into an integration over energy expresses the
fractional energy loss into ions and electrons as
\begin{eqnarray}
\frac{E_\smI}{E_0} &=& 
\int_0^{E_0} \frac{dE}{E_0} \, \frac{ {\cal A}_\smI(E)}{{\cal A}(E) }
- \int_0^\infty \frac{dE}{E_0} \, \frac{ {\cal A}_\smI(E)}{{\cal A}(E) }
  \,   \frac{2 \, \beta^{3/2}}{\sqrt\pi} \, 
      \int_E^\infty dE'  \, \sqrt{E'} \,  e^{-\beta E'} 
\label{Ianswer}
\end{eqnarray}
and
\begin{eqnarray}
\frac{E_e}{E_0} &=& 
\int_0^{E_0} \frac{dE}{E_0} \, \frac{ {\cal A}_e(E)}{{\cal A}(E) }
- \int_0^\infty \frac{dE}{E_0} \, \frac{ {\cal A}_e(E)}{{\cal A}(E) }
  \,   \frac{2 \, \beta^{3/2}}{\sqrt\pi} \, 
      \int_E^\infty dE'  \, \sqrt{E'} \,  e^{-\beta E'} \,,
\label{eanswer}
\end{eqnarray}
where $\beta = 1/T$. Adding these equations gives 
\begin{eqnarray}
 \left[ E_\smI + E_e \right] &=& 
  E_0 - \int_0^\infty dE \, 
  \,   \frac{2 \, \beta^{3/2}}{\sqrt\pi} \, 
      \int_E^\infty dE'  \, \sqrt{E'} \,  e^{-\beta E'}
\nonumber\\
&=& 
  E_0 -  \frac{2 \, \beta^{3/2}}{\sqrt\pi} \, 
      \int_0^\infty dE'  \, {E'}^{3/2} \,  e^{-\beta E'}
\nonumber\\
&=&
  E_0 - \frac{3}{2} \, T \,,
\label{Esum}
\end{eqnarray}
where the second line follows from a partial integration, and the
last line from the definition of $\Gamma(5/2)$. This is just the 
obvious result of energy conservation previously stated in 
Eq.~(\ref{obvious}). 

The results (\ref{Ianswer}) and (\ref{eanswer}) can be simplified
for their explicit evaluation.  Writing these results with a trivial 
rearrangement of the terms presents them as:
\begin{eqnarray}
\frac{E_\smI}{E_0} &=& 
\int_0^{E_0} \frac{dE}{E_0} \, \frac{ {\cal A}_\smI(E)}{{\cal A}(E) }
\left[ 1 -
  \frac{2 \, \beta^{3/2}}{\sqrt\pi} \, 
      \int_E^\infty dE'  \, \sqrt{E'} \,  e^{-\beta E'} \right]
\nonumber\\
&& \qquad\qquad\qquad 
 - \int_{E_0}^\infty \frac{dE}{E_0} \, \frac{ {\cal A}_\smI(E)}{{\cal A}(E) }
  \,   \frac{2 \, \beta^{3/2}}{\sqrt\pi} \, 
      \int_E^\infty dE'  \, \sqrt{E'} \,  e^{-\beta E'} \,,
\label{IIanswer}
\end{eqnarray}
and
\begin{eqnarray}
\frac{E_e}{E_0} &=& 
\int_0^{E_0} \frac{dE}{E_0} \, \frac{ {\cal A}_e(E)}{{\cal A}(E) }
\left[ 1 -
  \frac{2 \, \beta^{3/2}}{\sqrt\pi} \, 
      \int_E^\infty dE'  \, \sqrt{E'} \,  e^{-\beta E'} \right]
\nonumber\\
&& \qquad\qquad\qquad 
 - \int_{E_0}^\infty \frac{dE}{E_0} \, \frac{ {\cal A}_e(E)}{{\cal A}(E) }
  \,   \frac{2 \, \beta^{3/2}}{\sqrt\pi} \, 
      \int_E^\infty dE'  \, \sqrt{E'} \,  e^{-\beta E'} \,.
\label{eeanswer}
\end{eqnarray}
As we shall see,  the second set of double integrals in 
Eqs.~(\ref{IIanswer}) and (\ref{eeanswer}) are exponentially small.
Hence it suffices to use the simple bounds
\begin{equation}
\frac{ {\cal A}_\smI(E)}{{\cal A}(E) } =
\frac{ {\cal A}_\smI(E)}{{\cal A}_\smI(E) + {\cal A}_e(E) } \leq 1 \,,
\end{equation}
and similarly
\begin{equation}
\frac{ {\cal A}_e(E)}{{\cal A}(E) } \leq 1 \,.
\end{equation}
Using these bounds, we encounter
\begin{eqnarray}
 - \int_{E_0}^\infty \frac{dE}{E_0}  \, \frac{2 \, \beta^{3/2}}{\sqrt\pi} \, 
      \int_E^\infty dE'  \, \sqrt{E'} \,  e^{-\beta E'} 
&=& - \frac{2 \,\beta^{3/2}}{\sqrt\pi} \, 
    \int_{E_0}^\infty dE'  \, \sqrt{E'} \,  e^{-\beta E'} 
\int_{E_0}^{E'} \frac{dE}{E_0}
\nonumber\\
&=&
 - \frac{2 \,\beta^{3/2}}{\sqrt\pi} \, 
    \int_{E_0}^\infty dE'  \, \sqrt{E'} \,  e^{-\beta E'} 
\left[ \frac{E'}{E_0} - 1 \right] \,.
\end{eqnarray}
The variable change $ E' = E_0 \, (x+1) $ presents this as
\begin{eqnarray}
 - \frac{2 \,\beta^{3/2}}{\sqrt\pi} \, E_0^{3/2} \, e^{-\beta E_0} \,
     \int_0^\infty dx \, (1+x)^{1/2} \, x \, e^{-\beta E_0 x}
\simeq - \frac{2}{\sqrt\pi} \, \frac{1}{\sqrt{\beta E_0}} \, 
          e^{-\beta E_0} \,,
\end{eqnarray}
with the evaluation on the right-hand side following from the fact
that $\beta E_0 \gg 1$ so that only small $x$ regions contribute
justifying the replacement $(1+x)^{1/2} \to 1$. Hence we indeed find that
the additional double integrals in the energy fractions
(\ref{IIanswer}) and (\ref{eeanswer}) are exponentially small, and so 
with very good accuracy we may write these fractions as
\begin{eqnarray}
\frac{E_\smI}{E_0} &=& 
\int_0^{E_0} \frac{dE}{E_0} \, \frac{ {\cal A}_\smI(E)}{{\cal A}(E) }
\left[ 1 -
  \frac{2 \, \beta^{3/2}}{\sqrt\pi} \, 
      \int_E^\infty dE'  \, \sqrt{E'} \,  e^{-\beta E'} \right] \,,
\label{Ianswerr}
\end{eqnarray}
and
\begin{eqnarray}
\frac{E_e}{E_0} &=& 
\int_0^{E_0} \frac{dE}{E_0} \, \frac{ {\cal A}_e(E)}{{\cal A}(E) }
\left[ 1 -
  \frac{2 \, \beta^{3/2}}{\sqrt\pi} \, 
      \int_E^\infty dE'  \, \sqrt{E'} \,  e^{-\beta E'} \right] \,.
\label{eanswerr}
\end{eqnarray}

Since
\begin{equation}
  \frac{2 \, \beta^{3/2}}{\sqrt\pi} \, 
    \int_0^\infty dE'  \, \sqrt{E'} \,  e^{-\beta E'}  = 1 \,, 
\label{unit}
\end{equation}
we may write
\begin{equation}
1- \frac{2 \, \beta^{3/2}}{\sqrt\pi} \, 
\int_E^\infty dE'  \, \sqrt{E'} \,  e^{-\beta E'} =
\frac{2 \, \beta^{3/2}}{\sqrt\pi} \, 
\int_0^E dE'  \, \sqrt{E'} \,  e^{-\beta E'} \,,
\end{equation} 
while partial integration gives
\begin{equation}
 \frac{2 \, \beta^{3/2}}{\sqrt\pi} \, 
\int_0^E dE'  \, \sqrt{E'} \,  e^{-\beta E'} =
- \sqrt{ \frac{ 4\beta  E }{\pi} }  \,  e^{-\beta E} +
 \sqrt{ \frac{ \beta}{\pi} }  \, \int_0^E dE'  \, 
            \frac{1}{\sqrt{E'}}  \,  e^{-\beta E'} \,.
\end{equation} 
Hence, using the definition 
\begin{equation}
{\rm erf}(x) = \frac{2}{\sqrt\pi} \, \int_0^x dy \, e^{-y^2} 
\label{erf}
\end{equation}
of the error function, we may write the results (\ref{Ianswerr}) and
(\ref{eanswerr}) as
\begin{eqnarray}
\frac{E_\smI}{E_0} &=& 
\int_0^{E_0} \frac{dE}{E_0} \, \frac{ {\cal A}_\smI(E)}{{\cal A}(E) }
\left[ {\rm erf}(\sqrt{\beta E} ) -  
\sqrt{ \frac{ 4\beta  E }{\pi} }  \,  e^{-\beta E} \right] \,,
\label{IIIanswer}
\end{eqnarray}
and
\begin{eqnarray}
\frac{E_e}{E_0} &=& 
\int_0^{E_0} \frac{dE}{E_0} \, \frac{ {\cal A}_e(E)}{{\cal A}(E) }
\left[ {\rm erf}(\sqrt{\beta E} ) -  
\sqrt{ \frac{ 4\beta  E }{\pi} }  \,  e^{-\beta E} \right] \,.
\label{eeeanswer}
\end{eqnarray}
These are the results quoted in Eqs.~(\ref{IIIIanswer}) and
(\ref{eeeeanswer}) in the Introduction.

\subsection{Differing Electron and Ion Temperatures}
\label{differing}

As we have seen, when the ion and electron temperatures of the
background plasma differ, $T_\smI \neq T_e$, both the physical
interpretation is richer and the mathematics becomes more
difficult. With different temperatures, there is the additional
physical process in which the final distribution of stopped
injected impurity particles works to bring the electrons and ions
into thermal equilibrium at a common temperature $T=T_\smI =
T_e$.  Moreover, mathematically, we must now work with Eq.~(\ref{barg}).

We use Eq.~(\ref{barg}) to return to the $\bar f(E)$ function, and 
insert the result for $\bar f(E)$ into Eqs.~(\ref{Ifrac}) and
(\ref{efrac}) to compute $E_\smI/E_0$ and $E_e/E_0$. 
To simplify the resulting formulae, and 
place them in a form that parallels those for the previous equal
ion-electron temperature case we note that 
\begin{eqnarray}
\left\{
\begin{array}{c}
{\cal A}_\smI(E) \\
{\cal A}_e(E)
\end{array}
\right\}
\left[ 1 + \left\{
\begin{array}{c}
T_\smI \\
T_e
\end{array}
\right\} 
 \frac{d}{dE} \right] \, e^{-S(E)} 
= 
\left\{
\begin{array}{c}
+ \\
-
\end{array}
\right\}
 \left[ T_e - T_\smI \right]
\frac{{\cal A}_\smI(E) {\cal A}_e(E)} {\langle T {\cal A}(E) \rangle} 
\,  e^{-S(E)} \,.
\end{eqnarray}
Hence, with the definition
\begin{eqnarray}
G(T_\smI,T_e;E_0) &=&
\int_0^\infty dE \, E \,  
\frac{{\cal A}_\smI(E) {\cal A}_e(E)} {\langle T {\cal A}(E) \rangle} 
\,  e^{-S(E)} 
\nonumber\\
&& \!\!\!\!\!
\int_0^E \frac{dE'}{E'} \,
\frac{e^{ + S(E') }}{ \langle T{\cal A}(E') \rangle } \,
 \left\{  \theta\left( E_0 - E' \right) \, 
 -    \int_{E'}^\infty dE'' \sqrt{E''} \, 
  \overline{\cal N} \, e^{-S(E'')}  \right\} \,,
\nonumber\\
&&
\label{GGdef}
\end{eqnarray}
the energy loss fractions may be expressed as
\begin{eqnarray}
\frac{E_\smI}{E_0} &=& \left[\frac{ T_e - T_\smI}{E_0} \right] \, 
G(T_\smI,T_e;E_0) 
\nonumber\\
&&
+  \int_0^{\infty} \frac{dE}{E_0} \,  \frac{T_\smI \, {\cal A}_\smI(E)}
{ \langle T{\cal A}(E') \rangle } \,
 \left\{  \theta\left( E_0 - E \right) \, 
 -    \int_{E}^\infty dE' \sqrt{E'} \,
  \overline{\cal N} \, e^{-S(E')}  \right\} \,,
\nonumber\\
&&
\label{IIIfrac}
\end{eqnarray}
and
\begin{eqnarray}
\frac{E_e}{E_0} &=& \left[\frac{ T_\smI - T_e}{E_0} \right] \, 
\, G(T_\smI,T_e;E_0) 
\nonumber\\
&&
+  \int_0^{\infty} \frac{dE}{E_0} \,  \frac{T_e \, {\cal A}_e(E)}
{ \langle T{\cal A}(E') \rangle }   \,
 \left\{  \theta\left( E_0 - E \right) \, 
 -    \int_{E}^\infty dE' \sqrt{E'} \,
  \overline{\cal N} \, e^{-S(E')}  \right\} \,.
\nonumber\\
&&
\label{eeefrac}
\end{eqnarray}

The second lines in the results (\ref{IIIfrac}) and (\ref{eeefrac}) are
straightforward generalizations of the common ion and electron temperature
forms (\ref{Ianswer}) and (\ref{eanswer}). The first lines of the new
results (\ref{IIIfrac}) and (\ref{eeefrac}) cancel when they are summed,
so that 
\begin{eqnarray}
E_\smI  + E_e
&=& 
+ \int_0^{E_0} dE 
-  \overline{\cal N} \, \int_0^\infty dE \, 
\int_{E}^\infty dE'  \, \sqrt{ E'} \,  e^{-S(E')} \,.
\end{eqnarray}
Upon interchanging integrals,
\begin{eqnarray}
  \int_0^\infty dE \, 
\int_{E}^\infty dE'  \, \sqrt{E'} \,  e^{-S(E')} &=&
\int_0^\infty dE  \, \sqrt{E} \,  e^{-S(E)} 
\, \int_0^E dE' 
\nonumber\\
&=&
\int_0^\infty dE  \, E \, \sqrt{E} \,  e^{-S(E)} \,.
\end{eqnarray}
Hence, on passing from an integration over energy to an equivalent momentum
integral and reverting to the corresponding normalization factor 
${\cal N}$, we have 
\begin{equation}
E_\smI  + E_e = E_0 
- 
\int \frac{d^3{\bf p}}{(2\pi\hbar)^3} \, E \, {\cal N} \, e^{-S(E)} 
\end{equation}
or, in view of Eq.~(\ref{Ebar}), 
\begin{equation}
E_\smI + E_e = E_0 - \bar E \,,
\label{total}
\end{equation}
in which $\bar E$ is the average energy to which an impurity particle 
relaxes.  This result is in accord with the previous Eq.~(\ref{obvious}).

The final energy integrals in Eqs.~(\ref{IIIfrac}) and
(\ref{eeefrac}) run from $E=0$ to $E \to \infty$. In each case, the final 
integration region involves the exponentially small factor
$\exp\{ - S(E_0) \} \simeq \exp\{ - E_0 / \bar T \} $,
where $\bar T$ is a typical plasma temperature.  This is a very small factor,
and hence this upper portion of the integration region may be safely
neglected to write the results as
\begin{eqnarray}
\frac{E_\smI}{E_0} &=& \left[\frac{ T_e - T_\smI}{E_0} \right] \, 
G(T_\smI,T_e;E_0) 
+  \int_0^{E_0} \frac{dE}{E_0} \,  \frac{T_\smI \, {\cal A}_\smI(E)}
{ \langle T{\cal A}(E') \rangle }   \,
 \left\{ 1
 -    \int_{E}^\infty dE' \sqrt{E'} \,
  \overline{\cal N} \, e^{-S(E')}  \right\} 
\nonumber\\
&&
\nonumber\\
&=&  \left[\frac{ T_e - T_\smI}{E_0} \right] \, 
G(T_\smI,T_e;E_0) 
+  \int_0^{E_0} \frac{dE}{E_0} \,  \frac{T_\smI \, {\cal A}_\smI(E)}
{ \langle T{\cal A}(E') \rangle }   \,
   \int_0^E dE' \sqrt{E'} \,
  \overline{\cal N} \, e^{-S(E')}  \,,
\label{IIIIfrac}
\end{eqnarray}
and
\begin{eqnarray}
\frac{E_e}{E_0} &=& \left[\frac{ T_\smI - T_e}{E_0} \right] \, 
\, G(T_\smI,T_e;E_0) 
+  \int_0^{E_0} \frac{dE}{E_0} \,  \frac{T_e \, {\cal A}_e(E)}
{ \langle T{\cal A}(E') \rangle }   \,
 \left\{  1
 -    \int_{E}^\infty dE' \sqrt{E'} \,
  \overline{\cal N} \, e^{-S(E')}  \right\} \,.
\nonumber\\
&&
\nonumber\\
&=& \left[\frac{ T_\smI - T_e}{E_0} \right] \, 
\, G(T_\smI,T_e;E_0) 
+  \int_0^{E_0} \frac{dE}{E_0} \,  \frac{T_e \, {\cal A}_e(E)}
{ \langle T{\cal A}(E') \rangle }   \,
 \int_0^E dE' \sqrt{E'} \,
  \overline{\cal N} \, e^{-S(E')}  \,.
\label{eeeefrac}
\end{eqnarray}
Here we have invoked the sum rule (\ref{summ}) to write the second
equalities above.

The work in Appendix \ref{hack} shows that the function $G$ can be
approximated, with an accuracy of a few percent, by 
\begin{eqnarray}
\!\! G(T_\smI,T_e;E_0) &=& 
\int_0^{E_0} dE \, E \,
 \frac{ {\cal A}_\smI(E) {\cal A}_e(E)}{\langle T {\cal A}(E)\rangle } 
  e^{-S(E)} \!
 \int_0^E \frac{dE'}{E'} 
\frac{e^{ + S(E') }}{ \langle T{\cal A}(E') \rangle } 
 \int_0^{E'} dE'' \sqrt{E''} \, 
  \overline{\cal N} \, e^{-S(E'')}
\nonumber\\
&& \qquad\qquad
   + \frac{{\cal A}_\smI(E_0) \, {\cal A}_e(E_0)}{{\cal A}^2(E_0)} \,.
\label{done}
\end{eqnarray}
Since the integration in the first line is over the finite interval
$(0, E_0)$ and since it involves only nested integrals, rather than a
three-dimensional integral with an arbitrary integrand that involves
an general function of three variables, its numerical evaluation is
not difficult.

The explicit forms for the ${\cal A}$ coefficients reviewed
in appendix \ref{app:acoeff} now enable the explicit computation of the energy
ratios $E_\smI/E_0$ that are presented in Fig.~\ref{fig:TiEbarFig3.eps} and
the tables of Appendix \ref{Tables}.

\section{Summary and Conclusion}
\label{Con}

We have developed a formalism that enables the calculations of the
energy fractions that a fast particle deposits to the ions and
electrons when it slows down in a plasma of ions and electrons that have
different temperatures. Such calculations have not be done
previously. Our work applies to background plasmas that are weakly to
moderately coupled --- the range of validity of this restriction was
discussed in the Introduction.

Since the background plasma is not in thermal equilibrium, a fast
particle ends in a ``schizophrenic'' distribution which we explicit
compute in Sec.~\ref{problem}.  As described in Sec.~\ref{exchange},
the final non-thermal distribution of the initial fast particles
provides a mechanism to bring the differing electron and ion temperatures
to a final common temperature, a process that now appears in addition
to the usual electron-ion relaxation interaction.

Although our general method applies to the slowing of any fast particle
in an arbitrary background plasma, we are specifically interested in 
DT nuclear fusion, and thus we present explicit numerical results for
an initial 3.54 Mev alpha slowing in an equimolar DT plasma.

For the case of equal ion and electron temperatures, the energy fractions
$E_\smI/E_0$ and $E_e/E_0$ that we compute are in agreement with previous
work to leading accuracy, but our results are more precise because we also
compute the exact coefficient of the first non-leading term which is 
proportional to the plasma density $n$. The comparison between our and 
previous results for the equal temperature case was discussed in the 
Introduction and shown there in Figs.~\ref{fig:frac26} and \ref{fig:change}.

In order to motivate and give the flavor of our results for the general case
in which the ions and electrons in the background plasma have different
temperatures, Fig.~\ref{fig:TiEbarFig3.eps} was presented in the Introduction.
The table that follows  gives detailed results for the energy
fractions $E_\smI/E_0$ and $E_e/E_0$ for an alpha particle with an initial
energy of 3.54 MeV slowing in equimolar DT plasmas of three different
densities and a variety of temperatures.  

\newpage

\begin{center}
{\bf DT FUSION ALPHA PARTICLE ENERGY DEPOSITED \\
INTO THE IONS FOR VARIOUS PLASMA CONDITIONS}
\end{center}

\label{Tables}

\null 

\vskip -1.3cm

\begin{table}[ht]
\caption{\footnoteskip
\qquad\qquad\quad 
 $E_\smI/E_0$ for a density $n_e=10^{24}\,{\rm cm}^{-3}$ over a range \\
\null \qquad\qquad\qquad\qquad  
 of electron and ion temperatures that are measured in keV. 
}
\begin{tabular}{|l||c|c|c|c|c|c|} \hline
   ~$T_\smI $  &  ~$T_e=10 $~ & ~$T_e=20 $~ &   ~$T_e=30 $~ &  
   ~$T_e=50$~  & ~$T_e=100 $~ & ~$T_e=200 $~   \\ \hline\hline
   ~10~ & ~0.248~& ~0.404~& ~0.513~ & ~0.660~ & ~0.834~& ~0.936~  \\
   ~30~ & ~0.234~& ~0.389~& ~0.497~ & ~0.644~ & ~0.821~& ~0.930~  \\
   ~50~ & ~0.220~& ~0.374~& ~0.481~ & ~0.628~ & ~0.807~& ~0.922~  \\
   ~100~& ~0.185~& ~0.336~& ~0.440~ & ~0.586~ & ~0.769~& ~0.892~  \\
   ~200~& ~0.126~& ~0.263~& ~0.361~ & ~0.502~ & ~0.689~& ~0.827~  \\
   ~300~& ~0.079~& ~0.197~& ~0.285~ & ~0.418~ & ~0.607~& ~0.760~ \\\hline
\end{tabular} 
\label{table:24}
\end{table}
\vskip -0.3cm
\begin{table}[h]
\caption{\footnoteskip 
\qquad\qquad\quad
  $E_\smI/E_0$ for a density $n_e=10^{25}\,{\rm cm}^{-3}$ over a range \\ 
\null \qquad\qquad\qquad\qquad
  of electron and ion temperatures that are measured in keV.
}
\begin{tabular}{|l||c|c|c|c|c|c|} \hline
   ~$T_\smI $  &  ~$T_e=10 $~ & ~$T_e=20 $~ &   ~$T_e=30 $~ &  
   ~$T_e=50$~  & ~$T_e=100 $~ & ~$T_e=200 $~   \\ \hline\hline
   ~10~ & ~0.267~& ~0.421~ & ~0.531~ & ~0.675~ & ~0.843~& ~0.939~  \\
   ~30~ & ~0.252~& ~0.406~ & ~0.515~ & ~0.659~ & ~0.830~& ~0.933~  \\
   ~50~ & ~0.236~& ~0.391~ & ~0.499~ & ~0.643~ & ~0.816~& ~0.925~  \\
   ~100~& ~0.200~& ~0.352~ & ~0.457~ & ~0.601~ & ~0.778~& ~0.895~  \\
   ~200~& ~0.139~& ~0.278~ & ~0.376~ & ~0.516~ & ~0.698~& ~0.831~  \\
   ~300~& ~0.089~& ~0.209~ & ~0.299~ & ~0.431~ & ~0.617~& ~0.764~ \\\hline
\end{tabular} 
\label{table:25}
\end{table}
\vskip -0.3cm
\begin{table}[b!]
\caption{\footnoteskip 
\qquad\qquad\quad
  $E_\smI/E_0$ for a density $n_e=10^{26}\,{\rm cm}^{-3}$ over  a range \\
\null \qquad\qquad\qquad\qquad
  of electron and ion temperatures that are measured in keV. 
}
\begin{tabular}{|l||c|c|c|c|c|c|} \hline
   ~$T_\smI $  &  ~$T_e=10 $~ & ~$T_e=20 $~ &   ~$T_e=30 $~ &  
   ~$T_e=50$~  & ~$T_e=100 $~ & ~$T_e=200 $~   \\ \hline\hline
   ~10~ & ~0.293~& ~0.446~ & ~0.555~ & ~0.694~ & ~0.854~& ~0.942~  \\
   ~30~ & ~0.276~& ~0.430~ & ~0.539~ & ~0.679~ & ~0.841~& ~0.936~  \\
   ~50~ & ~0.260~& ~0.415~ & ~0.523~ & ~0.663~ & ~0.827~& ~0.928~  \\
   ~100~& ~0.223~& ~0.375~ & ~0.481~ & ~0.620~ & ~0.789~& ~0.899~  \\
   ~200~& ~0.159~& ~0.299~ & ~0.398~ & ~0.534~ & ~0.710~& ~0.836~  \\
   ~300~& ~0.106~& ~0.228~ & ~0.318~ & ~0.449~ & ~0.630~& ~0.770~ \\\hline
\end{tabular} 
\label{table:26}
\end{table}

\newpage

\appendix

\section{The A-Coefficients}
\label{app:acoeff}

The Fokker-Planck equation described in the text involves two
scalar coefficient functions with only one of them, the ${\cal A}$ 
coefficient, entering into our problem of the partition of the energy loss
of a fast charged particle into the ions and electrons in the plasma.  The 
Fokker-Planck equation, and the coefficients ${\cal A}_\smI$ 
and ${\cal A}_e$ coming from the ions and electrons that are needed 
for our problem, were discussed extensively in BPS \cite{bps}.  There 
a method of dimensional continuation was employed to compute the 
${\cal A}_b$ which enables the short-distance, point Coulomb scattering 
to be joined with the long-distance, collective force in an unambiguous 
fashion that has no double counting.  This method was used to evaluate 
the ${\cal A}_b$ both to leading and to subleading order --- roughly 
speaking --- to order $n\ln n$ as well as $n$, where $n$ is the 
plasma number density (made dimensionless by the adduction of 
suitable parameters).  For completeness, we present here the results
of BPS.  Since their derivation is subtle, it cannot be sketched here. 

The coefficient for the interaction of an ``impurity particle'' of
energy $E$ or velocity $v_p$, ($E= m_p \, v_p^2 /2$)  with the 
species $b$ of the background plasma may conveniently be written as
\begin{equation}
  {\cal A}_b(v_p) =  
   {\cal A}^\smC_b(v_p) + {\cal A}^{\Delta Q}_b(v_p) \,,
\label{all}
\end{equation}
which is the same as Eq.~(10.25) of BPS, with
\begin{equation}
 {\cal A}^\smC_b(v_p) =
    {\cal A}^\smC_{b,\smS}(v_p) +  {\cal A}^\smLT_{b,\smR}(v_p) \,,
\end{equation}
which is the same as Eq.~(9.6) of BPS.
Here $ {\cal A}^\smC_b(v_p)$ has two terms. The first accounts for the 
hard Coulomb scattering in the classical limit, while the second accounts 
for the collective, long-distance effects, which are entirely classical.  
The  term ${\cal A}^{\Delta Q}_b(v_p) $ is the quantum-mechanical 
correction  to the scattering that vanishes in the limit in which Planck's
constant vanishes, $\hbar \to 0$.

The first classical piece is given by
\begin{eqnarray}
  {\cal A}^\smC_{b,\smS}(v_p)
  &=& 
  \frac{e_p^2\,\kappa^2_b}{4\pi} \,   
  \left( \frac{\beta_b m_b}{2\pi} \right)^{1/2}\!\!
  v_p\int_0^1 du \, u^{1/2} \,\exp\left\{ - \frac{1}{2} \,
  \beta_b m_b v^2_p \, u \right\}
\nonumber\\
  && \qquad
  \left[ -\ln \left(\beta_b  \frac{e_p e_b}{4\pi} \,
  K \, \frac{m_b}{m_{pb}} \, \frac{u}{1-u} \right) 
  - 2 \gamma + 2 \right] \,,
\label{wonderclassic}
\end{eqnarray}
which is contained in Eq.~(9.5) of BPS. The reduced mass $m_{pb}$ of the 
projectile ($p$) and plasma particle ($b$) is defined by 
\begin{equation}
\frac{1}{m_{pb}} = \frac{1}{m_p} + \frac{1}{m_b} \,.
\label{reduce}
\end{equation}
The second part of the classical contribution is given by 
\begin{eqnarray}
  {\cal A}^\smLT_{b,\smR}(v_p)
  &=&
  \frac{e_p^2}{4 \pi}\, \frac{i}{2 \pi}
  \int_{-1}^1 \! d\cos\theta\, \cos\theta\,
  \frac{\rho_b(v_p\cos\theta)}
  {\rho_{\rm total}(v_p\cos\theta)}\,F(v_p \cos\theta) 
  \ln\!\left\{\frac{F(v_p\cos\theta)}{K^2}\right\} \,, 
\label{nun}
\end{eqnarray}
which is contained in Eq.~(7.26) of BPS. Here $\rho_\text{total}(v)$ is 
the spectral weight,
\begin{eqnarray}
  \rho_\text{total}(v)
  &=&
  {\sum}_b\, \rho_b\!\left(v\right) \ ,
\end{eqnarray}
with
\begin{eqnarray}
  \rho_b(v) 
  = 
  \kappa_b^2\,\sqrt{\frac{\beta_b m_b}{2\pi}}\, v\,
  \exp\!\left\{-\frac{1}{2}\,\beta_b m_b\, v^2\right\} \,,
\label{rhototdef}
\end{eqnarray}
as introduced in BPS Eqs.~(7.9) and (7.10). With these definitions,
it is not hard to show (as is explicitly done in BPS) that the sum
${\cal A}^\smC_{b,\smS} +{\cal A}^\smLT_{b,\smR} $
is independent of the arbitrary  wave number $K$ that was
introduced for computational convenience. The function
$F(v_p \cos\theta)$ is related to the classical dielectric function
$\epsilon( k , k v_p \cos\theta )$ by
\begin{equation}
  k^2 \, \epsilon( k , k v_p \cos\theta ) =
  k^2 + F(v_p \cos\theta)  \,.
\label{struct}
\end{equation}
Here, consistent with our leading orders evaluation, the dielectric
function corresponds to the classical limit of the quantum ring sum.
Hence the complex-valued function $F(v)$ is defined by
\begin{eqnarray}
  F(v) 
  = 
 -  \int_{-\infty}^\infty \! du \, 
  \frac{\rho_\text{total}(u)}{v - u + i\eta} \,.
\label{disp}
\end{eqnarray}
Equations (\ref{struct}) and (\ref{disp}) are the formulae (7.7) 
and (7.8) of BPS.

The quantum correction is contained in Eq.~(10.27) of BPS, and it
reads
\begin{eqnarray}
  {\cal A}^{\Delta Q}_b(v_p)
  &=&
  -\frac{e_p^2\, \kappa_b^2}{4 \pi}\,
  \left( \frac{\beta_b m_b}{2\pi} \right)^{1/2}\,
  \frac{1}{2}\int_0^\infty dv_{pb}
  \bigg\{ 2\, {\rm Re} \, \psi \left( 1 + i \eta_{pb}
  \right) - \ln \eta^2_{pb}  \bigg\}
\nonumber\\
  &&
  \frac{1}{\beta_b m_b v_p v_{pb}} \, 
  \Bigg[ \exp\left\{ - \frac{1}{2}\, \beta_b
  m_b \left( v_p - v_{pb} \right)^2\right\} 
   \left( 1 - \frac{1}{\beta_b m_b v_p v_{pb} } \right) 
\nonumber\\
  && \qquad\qquad\quad
  + \exp\left\{ - \frac{1}{2} \beta_b m_b \left( v_p + v_{pb} 
  \right)^2\right\} 
   \left( 1 + \frac{1}{\beta_b m_b v_p v_{pb} } \right) 
\Bigg]	\,.
\label{regb}
\end{eqnarray}
Here $\psi(z) = d \ln\Gamma(z) / dz$ and, with the rationalized
Gaussian units that were used by BPS (and which we continue to use)
where the Coulomb potential energy between charges $e_a$ and $e_b$ a
distance $r_{ab}$ apart is given by $ V = e_a e_b / (4\pi \, r_{ab})$,
the formula contains the dimensionless quantum coupling
\begin{equation}
  \eta_{pb} = \frac{e_p e_b}{4\pi \hbar v_{pb} } \,.
\end{equation} 
\pagebreak[4]

The following figures illustrate the behavior of the 
${\cal A}$-coefficients for an equimolar DT plasma with an alpha 
particle projectile of kinetic energy $E$. Figures~\ref{fig:One}
and~\ref{fig:Two} plot the electron and ion components 
${\cal A}_e$, ${\cal A}_\smI$ and their sum 
${\cal A} = {\cal A}_e + {\cal A}_\smI$ for a plasma with 
electron number density $n_e= 1.0 \times 10^{25}\,{\rm cm}^{-3}$, 
electron temperature $T_e=10\,{\rm keV}$, and ion
temperatures of $T_\smI=10\,{\rm keV}$ and $100\,{\rm keV}$,
respectively.

\begin{figure}[h!]
\vskip-1cm
\includegraphics[scale=0.45]{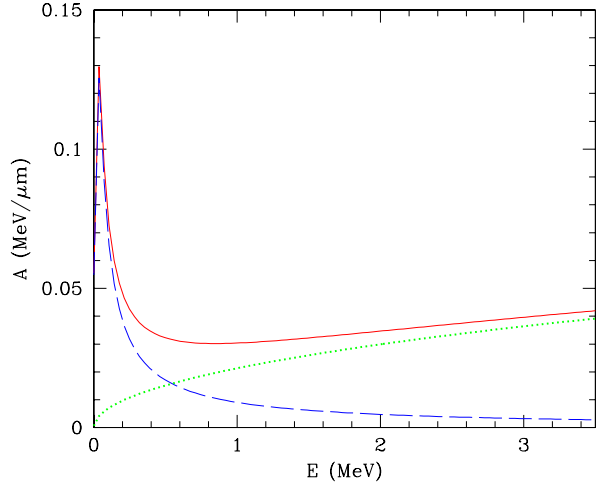}
\vskip-0.8cm
\caption{\captionskip The coefficients 
  ${\cal A}_\smI(E)$ ({dashed blue}), 
  ${\cal A}_e(E)$ ({dotted green}) 
  and 
  ${\cal A}(E)$ ({solid red}) as  
  functions of the kinetic energy $E$ of an $\alpha$ particle projectile. 
  The background plasma is equimolar DT with electron density 
  $n_e= 1.0 \times 10^{25}\,{\rm cm}^{-3}$ and electron-ion temperatures 
  $T_e=10\,{\rm keV}$ and $T_\smI=10\,{\rm keV}$. 
}
\label{fig:One}
\end{figure}
\begin{figure}[h!]
\vskip-1cm
\includegraphics[scale=0.45]{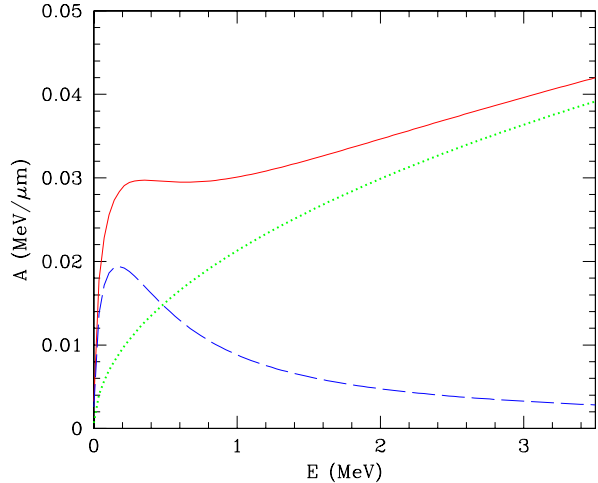}
\vskip-0.8cm
\caption{\captionskip
  As in Fig.~\ref{fig:One}, except $T_e=10\,{\rm keV}$ and 
  $T_\smI=100\,{\rm keV}$. The crossover energy $E = E_\smC$ 
  where ${\cal A}_e(E) = {\cal A}_\smI(E)$ is about the same in both 
  Figures; however, the peak value of the coefficient
  ${\cal A}_\smI$ is inversely proportional to~$T_\smI$. 
}
\label{fig:Two}
\end{figure}

\noindent
Figures \ref{fig:coeffte10ti10} and \ref{fig:coeffte10ti100}
illustrate the number density scaling of the ${\cal A}$-coefficients
by plotting ${\cal A}_e(E)/n_e$ and ${\cal A}_\smI(E)/n_e$, as a
function of the $\alpha$ particle energy $E$, over a wide range of
electron densities: $n_e =10^{25}$, $10^{26}$, and $10^{27}
\,\,{\rm cm}^{-3}$. As before, the electron temperature is
$T_e=10\,{\rm keV}$ and the ion temperatures are $T_\smI=10\,{\rm
keV}$ and $T_\smI = 100\,{\rm keV}$, respectively.

\begin{figure}[h!]
\vskip-1cm
\includegraphics[scale=0.45]{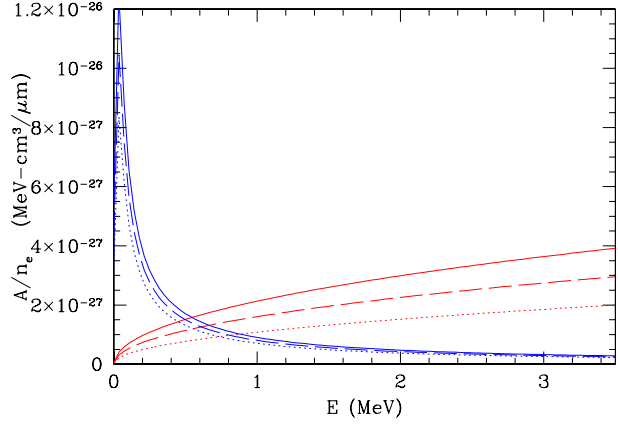}
\vskip-0.8cm 
\caption{\captionskip
  The $A$-coefficients for electrons and ions as a function of the
  $\alpha$ particle projectile energy $E$ in an equimolar DT plasma 
  with equal electron and ion temperatures, $T_e = T_\smI = 10$ keV.  
  The solid lines correspond to 
  $n_e = 1.0 \times 10^{25}\,{\rm cm}^{-3}$, 
  the dashed lines to $n_e = 1.0 \times 10^{26}\,{\rm cm}^{-3}$, and 
  the dotted lines to $ n_e = 1.0 \times 10^{27}\,{\rm cm}^{-3}$. 
  In each case, the $A$-coefficient has been rescaled by
  the corresponding number density $n_e$. The slowly rising (red) 
  curves are those for ${\cal A}_e$, while the sharply peaked 
  (blue) curves are for ${\cal A}_\smI$.
  }
\label{fig:coeffte10ti10}
\end{figure}

%
\begin{figure}[h!]
\vskip-1cm
\includegraphics[scale=0.45]{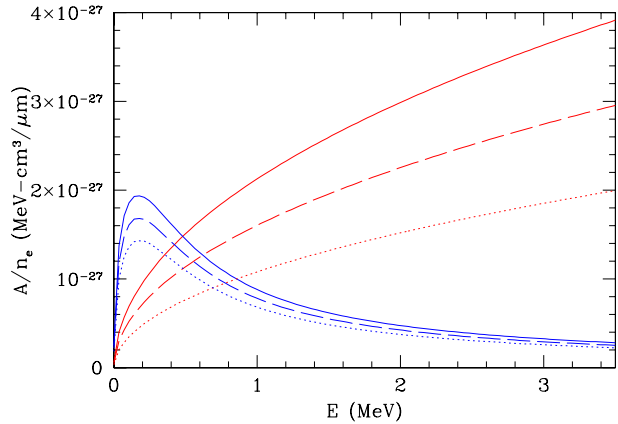}
\vskip-0.8cm 
\caption{\captionskip
  As in Fig.~\ref{fig:coeffte10ti10}, except with $T_e=10\,{\rm keV}$
  and $T_\smI=100\,{\rm keV}$. 
}
\label{fig:coeffte10ti100}
\end{figure}

\noindent
Because the ${\cal A}$-coefficients are proportional to the
Debye wave number squared, a quantity proportional to $n_e$, it is
no surprise that ${\cal A}_\smI$ and ${\cal A}_e$ approximately scale
with $n_e$. The Debye wave number also appears inside the logarithm
and the dielectric function, and for electrons this produces a much
more pronounced effect than for the much heavier ions: while ${\cal
A}_\smI/n_e$ is almost independent of $n_e$, the electron component
${\cal A}_e/n_e$ varies by a factor of two over the range of $n_e$.

\section{Asymptotic Limits}
\label{sec:limits}

We shall extract the large and small energy limits of the  
${\cal A}_b(v_p) $ function for the various plasma species $b$ 
from the general expressions in BPS\,\cite{bps}.  The energy is 
given by $E = m_p v_p^2 /2 $, where $m_p$ and $v_p$ are the mass and 
speed of the particle moving through the plasma, the projectile $p$.
We shall obtain the large and small limits of the projectile energy 
$E$ as compared to a typical plasma temperature $T$.

\subsection{$ {\bf E \ll T}$: Electrons and Ions}

In the low velocity limit, ${\cal A}_b(v_p) $ vanishes linearly with
$v_p$, and so we write
\begin{eqnarray}
v_p \to 0 \,: &&
\nonumber\\
&& {\cal A}_b(v_p) =  \frac{e_p^2\,\kappa_b^2}{4\pi} \,
    \left( \frac{\beta_b m_b}{2\pi} \right)^{1/2}v_p \, 
      \left\{ A_b^\smC + A_b^{\Delta Q} \right\} \,,
\label{low}
\end{eqnarray}
with two constants  $A_b^\smC$ and $A_b^{\Delta Q}$.  These two constants 
arise from the low velocity limit of the classical and quantum
pieces of Eq.~(\ref{all}). The classical piece has already been
calculated by BPS, where it is contained in their Eq.~(9.9), so 
there is no need to do it here.  The result is:
\begin{equation}
A_b^\smC = \frac{2}{3} \left[ \ln \left( 
     \frac{16\pi}{e_p e_b \, \beta_b \kappa_\smD} \, \frac{m_{pb}}{m_b}
       \right ) - \frac{1}{2} - 2 \gamma \right] \,,
\label{classvzero}
\end{equation}
in which $m_{pb}$ is the reduced mass defined in Eq.~(\ref{reduce})
in the previous Appendix, and $\gamma = 0.577 \dots $ is Euler's 
constant, and
\begin{equation}
\kappa_\smD^2 = {\sum}_b \, \kappa_b^2 
              = {\sum}_b \, \beta_b \, e_b^2 \, n_b \,.
\end{equation}
To bring out the size of this classical part, we define
a plasma coupling by
\begin{equation}
g_{pb} = \frac{e_p e_b \beta_b \kappa_\smD}{4\pi} 
       = \frac{e_pe_b}{4\pi \lambda_\smD} \, \frac{1}{T_b} \,,
\end{equation}
in which $\lambda_\smD = 1 / \kappa_\smD$ is the Debye length and 
$T_b = 1 / \beta_b$ is the temperature of plasma species $b$. Then
we may write
\begin{equation}
A_b^\smC = \frac{2}{3} \left[ \ln \left( 
     \frac{4}{g_{pb}} \, \frac{m_{pb}}{m_b}
       \right ) - \frac{1}{2} - 2 \gamma \right] \,,
\label{classvzeroo}
\end{equation}
which shows that $A_b^\smC > 0$, since $g_{pb}$ must be small for our
perturbative computation to hold.  Note that when the electron and ion
temperatures are not vastly different, the ions dominate in the low
velocity limit (\ref{low}) by a factor $(m_i/m_e)^{1/2}$. Moreover,
since $\kappa^2 \sim 1 / T $, this ionic contribution to the ${\cal
A}$-coefficient has the temperature factor $T_\smI^{-3/2}$ and thus
increases as the ion temperature is lowered. The corrections to the low
energy limit (\ref{low}) are of relative order $E/T$.

The low velocity limit of the quantum correction Eq.~(\ref{regb}) above
was not previously calculated in BPS because there the low velocity limit of 
$dE/dx$ was used only to compare with a computer simulation involving 
classical dynamics, and therefore the quantum correction was not needed.
The needed quantum part is contained in Eq.~(10.27) of BPS which provides 
the limit 
\begin{eqnarray}
&& v_p \to 0 \,:
\nonumber\\
&& \qquad\qquad
A_b^{\Delta Q} = - \frac{1}{3} \beta _b m_b 
\int_0^\infty dv_{pb} \, v_{pb} \, 
\exp\left\{ - \frac{1}{2} \, \beta_b m_b v_{pb}^2 \right\}
  \left[2 \, {\rm Re} \, \psi\left( 1 + i \eta_{pb} \right) 
        - \ln \eta^2_{pb} \right] \,.
\nonumber\\
&&
\label{DQ}
\end{eqnarray}
To bring out the character of  Eq.~(\ref{DQ}), we introduce a thermal 
velocity $\bar v_b$ by
\begin{equation}
\frac{1}{2} m_b \bar v_b^2 = \frac{3}{2} \, T_b \,, 
\label{barv}
\end{equation}
or
\begin{equation}
 \bar v_b^2 = \frac{3}{\beta_b m_b} \,,
\end{equation}
and a corresponding quantum parameter
\begin{equation} 
\bar\eta_{pb} = \frac{e_p e_b}{4\pi \hbar \bar v_b} \,.
\end{equation}
We then change the integration variable,
\begin{equation}
v_{pb} = \frac{e_p e_b}{4\pi\hbar \, \eta_{pb}} 
       = \frac{e_p e_b}{4\pi \hbar} \, u 
       = \bar\eta_{pb} \, \bar v_b \, u \,,
\end{equation}
to obtain
\begin{equation}
A_b^{\Delta Q} = A_b^{\Delta Q}(\bar\eta_{pb}) = 
  - \bar\eta_{pb}^2 \,
\int_0^\infty du \, u \, 
           \exp\left\{ - \frac{3}{2} \bar\eta_{pb}^2 \, u^2 \right\}
  \left[2 \,  {\rm Re} \, \psi\left( 1 + \frac{i}{u} \right) 
        + \ln u^2 \right] \,.
\label{Dq}
\end{equation}

If we introduce the Bohr radius $a_0 = 4\pi \hbar^2 / e^2 m_e $ and use
the average squared thermal velocity definition (\ref{barv}), 
we can write
\begin{equation}
{\bar\eta_{pb}}^2 = \frac{1}{3} \, \left( \frac{e_p e_b}{e^2} \right)^2
\, \frac{m_b}{m_e} \, \frac{1}{T_b} \, \frac{e^2}{4\pi a_0} \simeq
\frac{1}{3} \, \left( \frac{e_p e_b}{e^2} \right)^2
\, \frac{m_b}{m_e} \, \frac{27 {\rm eV}}{T_b} \,.
\end{equation}
Thus for the charge and mass of a typical projectile particle such as
an alpha particle and for a typical hot plasma, we see that for the
electrons in the plasma $ {\bar\eta_{pe}}^2 \ll 1 $, while for the
ions in the plasma $ {\bar\eta_{pi}}^2 \gg 1 $ unless the ion
temperature is somewhat larger than 10 keV.

For $ {\bar\eta_{pe}}^2 \ll 1 $, the exponential does not rapidly damp
large $u$ values, and so the relevant piece of the integrand is that
with $u \gg 1$ where
\begin{equation}
\psi\left( 1 + \frac{i}{u} \right) \simeq \psi(1) = - \gamma \,,
\end{equation}
leading to
\begin{eqnarray}
A_e^{\Delta Q}(\bar\eta_{pe})
 &\simeq&  - \bar\eta_{pe}^2 \,
\int_0^\infty du \, u \, 
           \exp\left\{ - \frac{3}{2} \bar\eta_{pe}^2 \, u^2 \right\}
  \left[- 2 \, \gamma  + \ln u^2 \right] 
\nonumber\\
&=&  \frac{1}{3} \, \ln\left( \frac{3}{2} \,\bar\eta^2_{pe} \right)
 + \gamma \,.
\end{eqnarray}
Adding this result to the classical limit (\ref{classvzero}) gives
the complete plasma electron contribution for a low energy projectile:
\begin{eqnarray}
E \ll T \, &&  {\bar\eta_{pe}}^2 \ll 1 \,:
\nonumber\\
{\cal A}_e(v_p) &=& \frac{e_p^2 \, \kappa_e^2 }{4\pi} \, \left(
\frac{\beta_e m_e}{2\pi} \right)^{1/2} \, \frac{v_p}{3} \left[ \ln
  \left(\frac{8 T_e m_{pe}^2}{m_e \hbar^2 \, \kappa_\smD^2} \right) -
  \gamma -1 \right] \,.
\label{Lowe}
\end{eqnarray}

For $ {\bar\eta_{pi}}^2 \gg 1 $, the exponential rapidly damps
large $u$ values, and so the relevant piece of the integrand is that
with $u \ll 1$ where
\begin{equation}
  \left[2 \,  {\rm Re} \, \psi\left( 1 + \frac{i}{u} \right) 
        + \ln u^2 \right] \simeq \frac{1}{6} \, u^2 \,,
\end{equation}
and thus
\begin{eqnarray}
{\bar\eta_{pi}}^2 \gg 1 \,: &&
\nonumber\\
&&
A_i^{\Delta Q}(\bar\eta_{pi}) \simeq
- \frac{\bar\eta_{pi}^2}{6} \,
\int_0^\infty du \, u^3 \, 
           \exp\left\{ - \frac{3}{2} \bar\eta_{pi}^2 \, u^2 \right\}
=  - \frac{1}{27} \, {\bar\eta}_{pi}^{\, -2} \,.
\end{eqnarray}
Since this is a very small correction to $A_i^\smC > 0$, it may be
neglected, and we may use the pure classical limit (\ref{classvzero})
for the ion contribution:
\begin{eqnarray}
   E \ll T \,, &&  {\bar\eta_{pi}}^2 \gg 1 \,:
\nonumber\\
  {\cal A}_i(v_p) &=&  \frac{e_p^2\,\kappa_i^2}{4\pi} \,
  \left( \frac{\beta_i m_i}{2\pi} \right)^{1/2} \, 
  \frac{2 v_p}{3} \left[ \ln \left( 
  \frac{16\pi}{e_p e_i \, \beta_i \kappa_\smD} \, \frac{m_{pi}}{m_i}
  \right ) - \frac{1}{2} - 2 \gamma \right] \,.
\label{LowI}
\end{eqnarray}
The total contribution of the ions in the plasma in this case is obviously
\begin{eqnarray}
E \ll T \,,  \quad {\bar\eta_{pi}}^2 \gg 1 \,: 
\qquad \qquad {\cal A}_\smI(v_p) = {\sum}_i {\cal A}_i(v_p) \,.
\label{LowIsum}
\end{eqnarray}
\begin{figure}[h!]
\vskip-1cm
\includegraphics[scale=0.45]{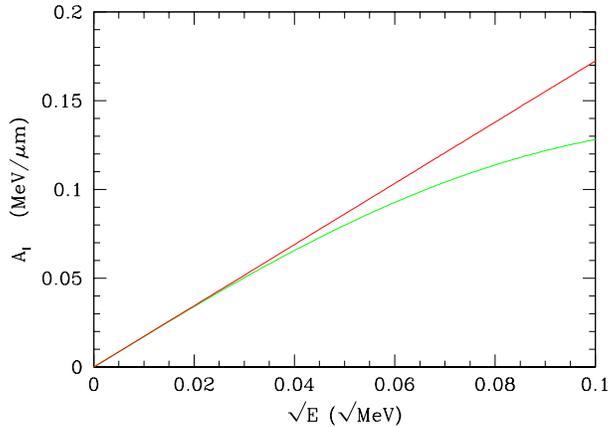}
\vskip-0.8cm 
\caption{\captionskip
  The ion contribution (green) plotted with the corresponding
  low-energy approximate form (red-linear) given by Eqs.~(\ref{LowI})
  and (\ref{LowIsum}). The plasma is equimolar DT with
  $T_e=T_\smI=10\,{\rm keV}$ and $n_e = 1.0 \times 10^{25}\,{\rm
  cm}^{-3}$, and the projectile is an $\alpha$ particle. For these
  parameters, the plasma coupling is $g_e=0.0006$. Since the
  leading-order small-energy behavior is proportional to $v_p$, the
  graph of ${\cal A}_\smI$ against $\sqrt{E}$ is linear in this
  region.  }
\label{fig:005.04}
\end{figure}
\begin{figure}[h!]
\vskip-1cm
\includegraphics[scale=0.45]{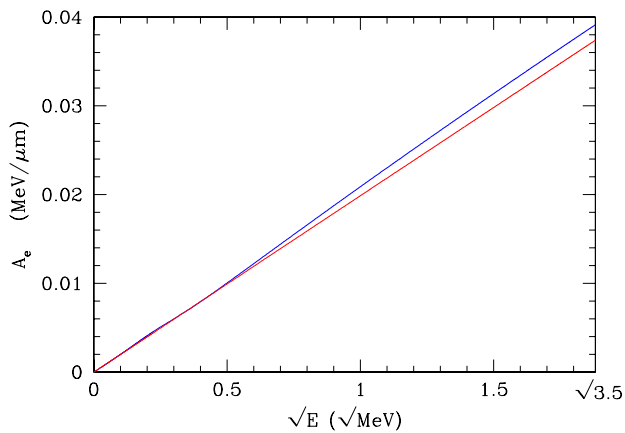}
\vskip-0.8cm 
\caption{\captionskip
  The electron contribution (blue) plotted with
  the corresponding low-energy approximate form (red-linear) given by
  Eq.~(\ref{Lowe}). The plasma is the same as in the
  previous figure. Note that 
  the linear approximation holds well into the $DT$ fusion production 
  energy of $3.54\,{\rm MeV}$ for the $\alpha$ particles. 
  }
\label{fig:006.04}
\end{figure}

In Figs.~\ref{fig:005.04} and \ref{fig:006.04} we plot the ion and
electron ${\cal A}$-coefficients for an equimolar $DT$ plasma with an 
electron density $n_e = 1.0 \times 10^{25} \, {\rm cm}^{-3}$ and equal
electron and ion temperatures $T_e = T_\smI = 10 \, {\rm keV}$ against 
the square root of the projectile energy $\sqrt{E}$. We make this choice
because in the small-energy  regime
the coefficients are linear in the projectile velocity; therefore, the
graphs exhibit linear behavior until they start to depart from the
low energy limit. Figures~\ref{fig:007.04} and
\ref{fig:008.04} plot the ${\cal A}$-coefficients  for an equimolar
DT plasma with $T_e=10\,{\rm keV}$ and $T_\smI=100\,{\rm keV}$ and with an
electron density $n_e = 1.0 \times 10^{25} \, {\rm cm}^{-3}$. 
\vskip -1cm
\begin{figure}[h!]
\vskip-1cm
\includegraphics[scale=0.45]{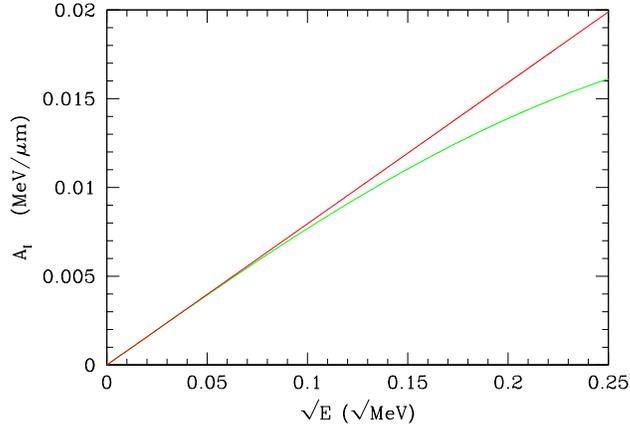}
\vskip-0.8cm 
\caption{\captionskip
  The ion contribution (green) plotted with the
  corresponding low-energy approximate form (red-linear) given by
  Eqs.~(\ref{LowI}) and (\ref{LowIsum}). The plasma is equimolar DT
  with $T_e=10\,{\rm keV}$, $T_\smI=100\,{\rm keV}$ and
  $n_e = 1.0 \times 10^{25}\,{\rm cm}^{-3}$, and the projectile is an 
  $\alpha$ particle. 
  }
\label{fig:007.04}
\end{figure}
\begin{figure}[h!]
\vskip-1cm
\includegraphics[scale=0.45]{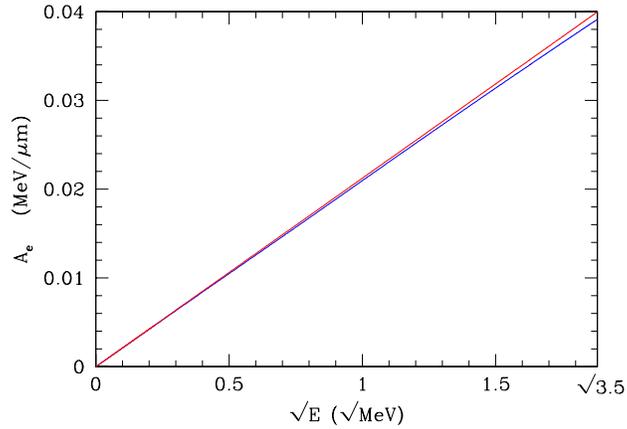}
\vskip-0.8cm 
\caption{\captionskip
  The electron contribution (blue) plotted with
  the corresponding low-energy approximate (red-linear) form (\ref{Lowe}). 
  The plasma is the same as in the previous figure. 
  }
\label{fig:008.04}
\end{figure}
\noindent
In Figs.~\ref{fig:009.04} and \ref{fig:010.04} the temperatures 
are changed to $T_e = 100 \, {\rm keV}$ and $T_\smI = 10 \, {\rm keV}$. 
\begin{figure}[h!]
\vskip-1cm
\includegraphics[scale=0.45]{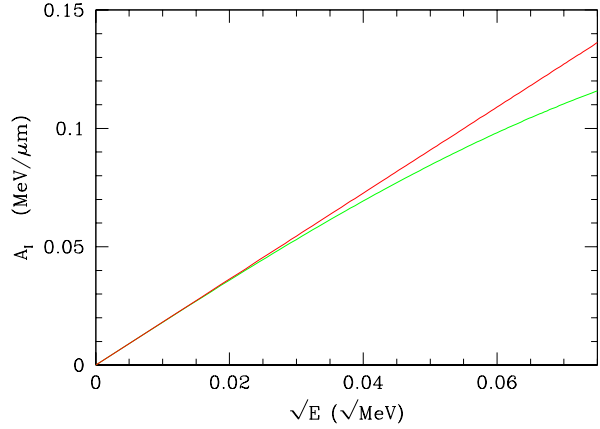}
\vskip-0.8cm 
\caption{\captionskip
  The ion contribution (green) plotted with the
  corresponding low-energy approximate (red-linear) form  
  (\ref{LowI}) and (\ref{LowIsum}). The plasma is equimolar DT
  with $T_e=100\,{\rm keV}$, $T_\smI=10\,{\rm keV}$ and
  $n_e = 1.0 \times 10^{25}\,{\rm cm}^{-3}$, and the projectile 
  is an $\alpha$ particle.
  } 
\label{fig:009.04}
\end{figure}
\begin{figure}[h!]
\vskip-1cm
\includegraphics[scale=0.45]{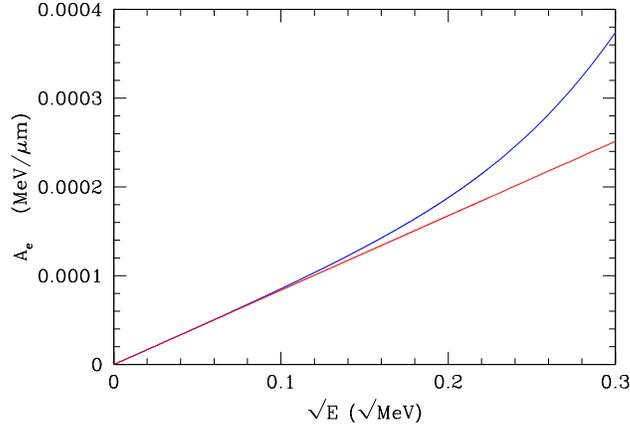}
\vskip-0.8cm 
\caption{\captionskip
  The electron contribution (blue) plotted with
  the corresponding low-energy approximate (red-linear) form 
  (\ref{Lowe}). The plasma is equimolar DT
  with $T_e=100\,{\rm keV}$, $T_\smI=10\,{\rm keV}$ and
  $n_e = 1.0 \times 10^{25}\,{\rm cm}^{-3}$, and the projectile 
  is an $\alpha$ particle. 
  }
\label{fig:010.04}
\end{figure}

\newpage

\noindent
In Fig.~\ref{fig:210.04} the coefficients ${\cal A}_e$ and 
${\cal A}_\smI$ are plotted together for three different 
temperatures.  
\begin{figure}[h!]
\vskip-1cm
\includegraphics[scale=0.45]{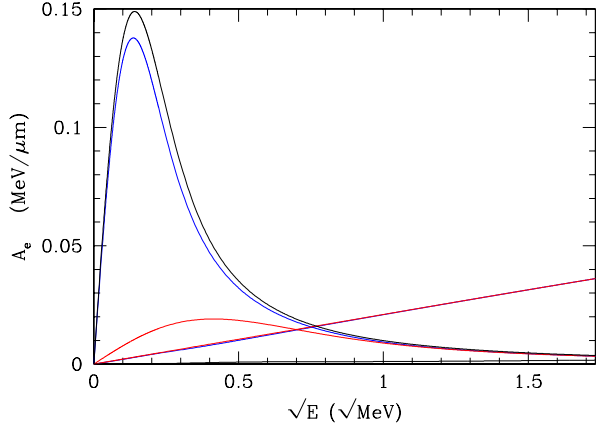}
\vskip-0.8cm 
\caption{\captionskip
  The coefficients $A_e$ and $A_\smI$ are plotted together for three
  different temperatures for an equimolar DT plasma with an electron
  number density $n_e = 1.0 \times 10^{25} \, {\rm cm}^{-3}$.  
  The ion contributions $A_\smI$ peak 
  to the left in the figure. The temperatures are: 
  (i) $T_e=10\,{\rm keV}$ and $T_\smI=10\,{\rm keV}$ (blue), 
  (ii) $T_e=10\,{\rm keV}$ and $T_\smI=100\,{\rm keV}$ (red),
  (iii) $T_e=100\,{\rm keV}$ and $T_\smI=10\,{\rm keV}$ (black).
  The electron contributions $A_e$ for cases (i) and (ii) are
  almost equal, whereas for case (iii) ${\cal A}_e$ is very small.
  }
\label{fig:210.04}
\end{figure}

\pagebreak
\subsection{$\bm{E \gg T}$: Total Ionic Contribution } 

For the total ionic contribution, it is convenient to first work out
the regular part of the long-distance, dielectric contribution because
it is the same for both cases of classical and quantum-mechanical
scattering.  With a trivial integration variable change,
Eq.~(\ref{nun}) presents this contribution as
\begin{equation}
  {\cal A}_{\smI,\smR}^\smLT(v_p) 
  = 
  \frac{e_p^2}{4\pi} \,\frac{1}{v_p^2} \, \frac{i}{2\pi} 
  \int_{-v_p}^{+v_p} dv \, v \, \frac{\rho_\smI(v)}{\rho_{\rm total}(v)}
   \, F(v) \, \ln \left( \frac{F(v)}{K^2} \right) \,,
\label{I}
\end{equation}
where we now write
\begin{equation}
\rho_\smI(v) = {\sum}_i \, \rho_i(v) \,,
\end{equation}
so that
\begin{equation}
\rho_{\rm total}(v) = \rho_e(v) + \rho_\smI(v) \,,
\end{equation}
with the weight functions $\rho$ given by Eq.~(\ref{rhototdef}).
Assuming that the charges of the ions do not differ greatly from the charge
of the electron, then $\kappa_e^2 / \kappa^2_\smI \simeq T_\smI / T_e$ 
and the integrand of Eq.~(\ref{I}) involves a factor that has the behavior
\begin{equation}
  \frac{\rho_\smI(v)}{\rho_{\rm total}(v) } = 
  \frac{1}{1 + \rho_e(v)/\rho_\smI(v) } \simeq  
  \frac{1}{ 1 + (m_e T_\smI^3 / m_\smI T_e^3)^{1/2} \, 
  \exp\left\{ m_\smI v^2/2T_\smI \right\} } \,,
\end{equation}
where $m_\smI$ is a typical ion mass.  Thus, defining a typical ionic
thermal velocity $v_\smT$ by
\begin{equation}
 m_\smI \, v_\smT^2 = T_\smI \,, 
\end{equation}
this factor remains unity up to the critical velocity $v_{\rm crit}$ 
defined by 
\begin{equation}
 v_{\rm crit}^2 =  v_\smT^2 \, 
    \ln \left( \frac{m_\smI \, T^3_e}{m_e \, T^3_\smI} \right) \,,
\label{vC}
\end{equation}
after which it falls fairly rapidly to zero. The logarithmic factor in 
Eq.~(\ref{vC}) is typically about a factor of 10.  So $v_{\rm crit}$ is 
somewhat larger than an ion thermal velocity yet it is considerably 
smaller than the electron thermal velocity.  

In this region in which the factor $\rho_\smI(v)/ \rho_{\rm total}(v)$ 
of the integrand is non-vanishing, the function  [Eq.~(\ref{disp}) above]
\begin{equation}
F(v) = - \int_{-\infty}^{+\infty} du  
\frac{\rho_{\rm total}(u)}{ v - u +i\eta} 
\end{equation}
has the form
\begin{equation}
F(v) = \tilde F(v) =  \kappa_e^2  + F_\smI(v) \,,
\label{Ftilde}
\end{equation}
where
\begin{equation}
F_\smI(v) = - \int_{-\infty}^{+\infty} du  
    \frac{\rho_\smI(u)}{ v - u +i\eta} \,.
\end{equation}
This is so because the velocity $v$, which must be less than
$v_{\rm crit}$, is much less than the electron thermal velocity.  
Hence the electron part of $F(v)$ takes on its low velocity limit,
the electron Debye wave number squared $\kappa_e^2$. We place 
the form (\ref{Ftilde}) into Eq.~(\ref{I}) to obtain
\begin{equation}
  {\cal A}_{\smI,\smR}^\smLT(v_p) 
 \simeq
  \frac{e_p^2}{4\pi} \,\frac{1}{v_p^2} \, \frac{i}{2\pi} 
  \int_{-v_p}^{+v_p} dv \, v \, \frac{\rho_\smI(v)}{\rho_{\rm total}(v)}
   \, \tilde F(v) \, 
   \ln \left( \frac{ \tilde F(v)}{\kappa_e^2} \right) \,.
\label{Itilde}
\end{equation}
Here we have replaced the arbitrary intermediate wave number $K$ by
the electron Debye wave number $\kappa_e$ because now 
\begin{eqnarray}
v \to \infty \,: \qquad\qquad
\frac{\tilde F(v)}{\kappa_e^2} \to 
                1 - \frac{\omega_\smI^2}{\kappa_e^2 \, v^2} \,,
\end{eqnarray}
where
\begin{equation}
\omega_\smI^2 = {\sum}_i \, \omega_i^2  
           = {\sum}_i \, \frac{e_i^2 \, n_i}{m_i} \,,
\end{equation}
and so
$ \ln ( \tilde F(v) / \kappa_e^2 ) $
vanishes for large $v$. 

In order of magnitude, 
\begin{equation}
\frac{\omega_\smI^2}{\kappa_e^2 \, v^2} \simeq \frac{T_e}{m_\smI \, v^2} 
= \frac{T_e}{T_\smI} \, \frac{v_\smT^2}{v^2} \,.
\end{equation}
Hence, since $v_\smT^2$ is much less than $v_{\rm crit}^2$, unless 
$T_e$ is considerably larger than $T_\smI$, the final factor 
in the integral (\ref{Itilde}), $ \ln ( \tilde F(v) / \kappa_e^2 ) $, 
vanishes before $\rho_\smI(v)/\rho_{\rm total}(v)$ departs 
significantly from unity. Hence we simply take 
$\rho_\smI(v)/\rho_{\rm total}(v) = 1$ and write
\begin{equation}
  {\cal A}_{\smI,\smR}^\smLT(v_p) 
 \simeq
  \frac{e_p^2}{4\pi} \,\frac{1}{v_p^2} \, \frac{i}{2\pi} 
  \int_{-v_p}^{+v_p} dv \, v \,  \tilde F(v) \, 
   \ln \left( \frac{ \tilde F(v)}{\kappa_e^2} \right) \,.
\end{equation}

The discussion above shows that when $v_p > v_{\rm crit}$, the 
limits of the integration may be replaced by $\pm \infty$. Recalling 
the definition (\ref{vC}) of the critical velocity $v_{\rm crit}$, and
assuming that the projectile mass $m_p$ is about the same as the typical
ion mass $m_\smI$ in the plasma, we can now state that  
\begin{eqnarray}
E >  T \, \ln \left( \frac{m_\smI T_e^3}{m_e T_\smI^3} \right) \,: &&
\nonumber\\
  {\cal A}_{\smI,\smR}^\smLT(v_p) 
  &=& 
  \frac{e_p^2}{4\pi} \,\frac{1}{v_p^2} \, \frac{i}{2\pi} 
  \int_{-\infty}^{+\infty} dv \, v \,  \tilde F(v) \, 
   \ln \left( \frac{ \tilde F(v)}{\kappa_e^2} \right) \,.
\end{eqnarray}
We should note the convergence of the integral requires that the 
integration  limits are to be taken in a rigorously symmetrical 
fashion with the integral performed between exactly $-v_p$ and 
$+v_p$ and then $v_p \to \infty$ taken. It is now a simple matter
to evaluate this limiting form.  Adding a semicircle in the upper
half plane of radius $v_p$ gives a closed contour integral with no
interior singularities that accordingly vanishes. Hence the value
of the original integral is the negative of the integral over this
large semicircle, an integral that is trivially performed using 
the limiting forms listed before. Thus  
\begin{eqnarray}
E >  T \, \ln \left( \frac{m_\smI T_e^3}{m_e T_\smI^3} \right) \,: \qquad
{\cal A}_{\smI,\smR}^\smLT(v_p) = - \frac{e_p^2}{4\pi} \, 
\frac{\omega_\smI^2 }{2 \, v_p^2}  \,.
\label{III}
\end{eqnarray}

With the long-distance, dielectric ionic contribution evaluated in the
projectile high-energy limit, we can now compute the complete function
${\cal A}_i(v_p)$ in this limit.  To do so, we must distinguish two
cases for the remaining hard scattering contribution.    

\subsubsection{$E \gg T$, $\eta_{pi}^{2} \gg 1$}

As shown in detail in Sec.~10 of BPS, the classical scattering 
contribution dominates when the Coulomb parameter $\eta_{pi}$ is large,
with the first quantum-mechanical correction of relative order
\begin{equation}
\eta_{pi}^{-2} = \left(\frac{4 \pi \hbar v_p}{e_p e_i}\right)^2
            = \left(\frac{e^2}{e_p e_i}\right)^2 \,
                \frac{2 E}{\alpha^2 \, m_p c^2} \,,
\end{equation}
where $\alpha \simeq 1/137$ is the fine structure constant. In this
classical limit, the scattering contribution is given by 
Eq.~(\ref{wonderclassic}).  For the previous evaluation of the
dielectric contribution to hold, we must choose $K = \kappa_e$ so
that this formula reads
\begin{eqnarray}
  {\cal A}^\smC_{i,\smS}(v_p)
  &=& 
  \frac{e_p^2 \, \kappa_i^2 }{4\pi}\, 
  \left( \frac{\beta_i m_i}{2\pi} \right)^{1/2}\!\!
  v_p\int_0^1 du \, u^{1/2} \,\exp\left\{ - \frac{1}{2}\,
  \beta_i m_i v^2_p \, u \right\}
\nonumber\\
  && \qquad
  \left[ -\ln \left(\beta_i  \frac{e_p e_i}{4\pi} \,
  \kappa_e \, \frac{m_i}{m_{pi}} \, \frac{u}{1-u} \right) 
  - 2 \gamma + 2 \right] \,.
\end{eqnarray}
Since $m_i v^2_p / 2 \sim E \gg T_i$, only small $u$ values are
significant. Hence we can approximate $1-u = 1$ within the logarithm
and extend the integration limit to $u \to \infty$.  With the variable
change $ (\beta_i m_i v_p^2/ 2) \, u = s^2 $, we obtain the
high-energy limit
\begin{eqnarray}
  {\cal A}^\smC_{i,\smS}(v_p)
  &=& 
  \frac{e_p^2}{4\pi}  \, \frac{\omega_i^2}{ v_p^2} 
\, \frac{4}{\sqrt\pi} \int_0^\infty ds \, s^2 \,\exp\left\{ - s^2 \right\}
\left[  
\ln\left(\frac{4\pi}{e_p e_i \kappa_e}\,\frac{m_{pi}v_p^2}{2}\right) 
- \ln s^2 - 2 \gamma + 2 \right] \,,
\nonumber\\
&&
\end{eqnarray}
where we have used $\kappa_i^2 / \beta_i m_i = \omega_i^2$. Here we have
the integrals
\begin{equation}
 \frac{4}{\sqrt\pi} \int_0^\infty ds \, s^2 \,\exp\left\{ - s^2 \right\}
= 1 \,,
\end{equation}
and
\begin{eqnarray}
 \frac{4}{\sqrt\pi} \int_0^\infty ds \, s^2 \,\exp\left\{ - s^2 \right\}
\, \ln s^2 &=& \frac{2}{\sqrt\pi} \, \Gamma\left(\frac{3}{2}\right)
\, \psi\left(\frac{3}{2}\right) = \psi\left(\frac{3}{2}\right) 
\nonumber\\
&=& 2 - \gamma  - \ln 4 \,.
\end{eqnarray}
Whence,
\begin{eqnarray}
  {\cal A}^\smC_{i,\smS}(v_p)
  &=& 
  \frac{e_p^2}{4\pi}  \, \frac{\omega_i^2}{ v_p^2} 
\,  \left\{
\ln\left(\frac{16\pi}{e_p e_i \kappa_e}\,\frac{m_{pi}v_p^2}{2}\right) 
 - \gamma \right\} \,,
\end{eqnarray}
which, summed over all the ions in the plasma and combined with the
previous long-distance result (\ref{III}) yields the total contribution
from the ions in the plasma:
\begin{eqnarray}
E \gg T \,, \quad \eta_{pi}^{2} &\gg& 1 \,:
\nonumber\\
  {\cal A}\smI(v_p) &=& {\sum}_i \, {\cal A}_i(v_p)
= {\sum}_i \, \left\{ {\cal A}^\smC_{i,\smS}(v_p) + {\cal
  A}^\smLT_{i,\smR}(v_p) \right\} 
\nonumber\\
  &=& 
  \frac{e_p^2}{4\pi}  \, \frac{1} {v_p^2} \, {\sum}_i \,
  \omega_i^2 \,
\,  \left\{ 
\ln\left(\frac{16\pi}{e_p e_i \kappa_e}\,\frac{m_{pi}v_p^2}{2}\right) 
 - \gamma - \frac{1}{2}  \right\} \,.
\label{IIII}
\end{eqnarray}
See Fig.~\ref{fig:014.04} for a comparison of $A_\smI$ with its asymptotic
form at large energy. 
\begin{figure}[ht!]
\vskip-1cm
\includegraphics[scale=0.45]{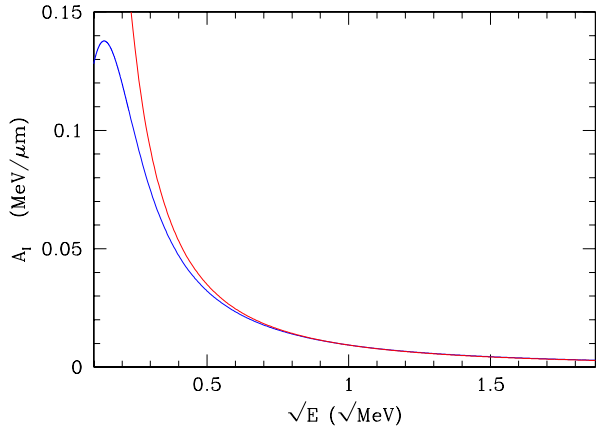}
\vskip-0.8cm 
\caption{\captionskip
  ${\cal A}_\smI$ (blue curve) {\em vs} its large-energy (red curve) 
  asymptotic form. The plasma is equimolar DT with 
  $n_e = 1.0 \times 10^{25} \, {\rm cm}^{-3}$ and 
  $ T_e = T_\smI = 10 \, {\rm keV}$.
  }
\label{fig:014.04}
\end{figure}

\subsubsection{$E \gg T$, $\eta_{pi}^2 \ll 1$}

In this case, we have the limit
\begin{eqnarray}
v_p \to \infty \,: &&
\nonumber\\
 &&
 {\cal A}^\smC_{i,\smS}(v_p) 
  + {\cal A}^{\Delta Q}_i(v_p) =
\frac{e_p^2}{4\pi} \, \frac{1}{v_p^2} \, {\sum}_i \, 
\omega_i^2  \ln \left( \frac{2 m_{pi} v_p}{\hbar\kappa_e} \right) 
\end{eqnarray}
which is contained in Eq.~(10.42) of BPS. 
Adding this result to Eq.~(\ref{III}) now provides the
complete $v_p \to \infty$ limit for the ion part of the ${\cal A}_\smI$
coefficient:
\begin{eqnarray}
E \gg T\,, \quad \eta_{pi}^2 \ll& 1 \,:
\nonumber\\
{\cal A}_\smI(v_p) &=&  \frac{e_p^2}{4\pi} \, \frac{1}{v_p^2} \, {\sum}_i \, 
\omega_i^2 \left\{ \ln \left( \frac{2 m_{pi} v_p}{\hbar\kappa_e} \right) 
            -  \frac{1}{2} \right\} \,. 
\label{V}
\end{eqnarray}

\subsubsection{Rough Estimate For Either Case}

The overall factor in the contribution of the ions to either of the
high velocity limits (\ref{IIII}) or (\ref{V}) of the ${\cal
A}$-coefficient has the same form as that of the electron contribution
(\ref{eeeeee}) given below, but with the major difference that the
squared electron plasma frequency $\omega_e^2$ is replaced by a sum of
squared ion plasma frequencies $\omega_i^2$, which are much smaller
than the electron contribution by the ratio $m_e / m_i $. For a rough
estimate of the size of the ionic contribution for high energy
projectiles, we approximate the logarithm in either limit (\ref{IIII})
or (\ref{V}) by a constant $L$ of order one, and approximate
$\omega_\smI^2 \simeq T_\smI \, \kappa_\smI / m_\smI $ and $ m_\smI
\simeq m_p $ to obtain
\begin{eqnarray}
E \gg T \,:  \qquad\qquad
{\cal A}_\smI(v_p) \simeq \frac{e_p^2 \, \kappa_\smI^2}{4 \pi} \,
         \frac{T_\smI}{E} \, L \,.
\label{Lion}
\end{eqnarray}

\subsection{$\bm{T \ll E \ll m_p T/m_e}$: Electronic Contribution } 

There is an intermediate range of projectile energies in which the projectile
energy is much larger that the temperature, $E \gg T$, but yet not so large
that we have  $  E \ll (m_p/ m_e) \,  T \sim 10^4 \, T$. We examine this range 
here.

We again need to work out its long-distance, dielectric contribution, and its
short-distance scattering contribution.

\subsubsection{Dielectric Part}

In the energy range specified, the typical velocity in the dielectric
function is small in comparison with the electron average thermal
velocity and large in comparison with an ion average thermal
velocity.  Hence, in this range
\begin{equation}
F(v) \simeq \kappa_e^2 - \frac{\omega_\smI^2}{v^2} + 
           \pi i \, \rho_{\rm total}(v) \,.
\end{equation}
Here, in the dominant integration range, 
\begin{equation}
\frac{\omega_\smI^2}{\kappa_e^2 \, v^2} \simeq \frac{T}{m_\smI \, v^2}
\ll 1 \,,
\end{equation}
and so we may simply write
\begin{equation}
F(v) \simeq \kappa_e^2 + \pi i \, \rho_{\rm total}(v) \,.
\label{simple}
\end{equation}
Moreover, in the dominant integration range, the imaginary part 
$ \pi \, \rho_{\rm total}(v) $
is small in comparison to $\kappa_e^2$.  Writing Eq.~(\ref{nun}) as 
\begin{eqnarray}
  {\cal A}^\smLT_{e,\smR}(v_p)
  &\simeq&
  \frac{e_p^2}{4 \pi}\, \frac{i}{2 \pi}
  \int_0^1 \! d\cos\theta\, \cos\theta \,
  \frac{\rho_e(v_p\cos\theta)}
  {\rho_{\rm total}(v_p\cos\theta)} \, \frac{1}{2} \,\Bigg\{ 
\nonumber\\
&& \qquad\qquad
\Big[ F(v_p \cos\theta) - F(-v_p \cos\theta) \Big] 
  \ln\!\left(\frac{F(v_p\cos\theta)F(-v_p \cos\theta)}{K^4}\right)
\nonumber\\
&& \qquad\qquad\qquad
+\Big[ F(v_p \cos\theta) + F(-v_p \cos\theta) \Big] 
  \ln\!\left(\frac{F(v_p\cos\theta)}{F(-v_p \cos\theta)}\right)
    \Bigg\} \,,
\end{eqnarray}
and using Eq.~(\ref{simple}) with the imaginary part treated to
first order, 
\begin{eqnarray}
  {\cal A}^\smLT_{e,\smR}(v_p)
  &\simeq&
 - \frac{e_p^2}{4 \pi}\, 
  \int_0^1 \! d\cos\theta\, \cos\theta \,
  \rho_e(v_p\cos\theta) \,\left\{
  \ln\!\left(\frac{\kappa_e^2}{K^2}\right) +1 \right\} \,.
\end{eqnarray}
In our energy range Eq.~(\ref{rhototdef}) becomes
\begin{eqnarray}
  \rho_e(v) 
  = 
  \kappa_e^2\,\sqrt{\frac{\beta_e m_e}{2\pi}}\, v \,,
\end{eqnarray}
and so
\begin{eqnarray}
  {\cal A}^\smLT_{e,\smR}(v_p)
  &\simeq&
  - \frac{e_p^2}{4 \pi}\, 
\kappa_e^2\,\sqrt{\frac{\beta_e m_e}{2\pi}}\, v_p \, \frac{1}{3} \, 
\left\{  \ln\!\left(\frac{\kappa_e^2}{K^2}\right) +1 \right\} \,.
\label{ediel}
\end{eqnarray}

\subsubsection{Scattering Part}

The electrons in the hot plasmas that we consider have such large
velocities that their scattering off the projectiles is quantum
mechanical.  This is described by Eq.~(10.41) of BPS  which gives
\begin{eqnarray}
&& \eta_{pe} \to 0 \,:
\nonumber\\
&& \qquad
 {\cal A}^\smC_{e,\smS}(v_p)  + {\cal A}^{\Delta Q}_e(v_p) =
   \frac{e_p^2\,\kappa^2_e}{4\pi}\, 
  \left( \frac{\beta_e m_e}{2\pi} \right)^{1/2}\!\!
  v_p\int_0^1 du \, u^{1/2} \,\exp\left\{ - \frac{1}{2}\,
  \beta_e m_e v^2_p \, u \right\}
\nonumber\\
  && \qquad\qquad\qquad\qquad\qquad\qquad\qquad
\frac{1}{2} \,  \left[ -\ln \left( 
\frac{\beta_e \hbar^2 K^2}{2 m_{pe}} \, 
\frac{m_e}{m_{pe}} \, \frac{u}{1-u} \right)   -  \gamma + 2 \right] \,.
\end{eqnarray}
With $ E = \frac{1}{2} \, m_p v_p^2 \ll m_p T / m_e $, the damping 
constant in the exponent $ \beta_e m_e v_p^2 / 2 $ is now small, 
not large as it was before. Hence the exponential may simply be replaced 
by unity, and we encounter the integrals
\begin{equation}
\int_0^1 du \, u^{1/2} = \frac{2}{3} \,,
\end{equation}
and
\begin{equation}
\int_0^1 du \, u^{1/2} \ln \left(\frac{u}{1-u}\right) = 
     \frac{2}{3} \, \left[ 2 - \ln 4 \right] \,.
\end{equation}
Hence
\begin{eqnarray}
 {\cal A}^\smC_{e,\smS}(v_p)  + {\cal A}^{\Delta Q}_e(v_p) 
&=&
   \frac{e_p^2\,\kappa^2_e}{4\pi}\, 
  \left( \frac{\beta_e m_e}{2\pi} \right)^{1/2}
  v_p \,\frac{1}{3} \, \left[
\ln \left( \frac{8 T_e m_{pe}^2}{m_e\hbar^2 K^2} \right) - \gamma \right] \,.
\label{escat}
\end{eqnarray}

\subsubsection{The Sum and a Rough Approximation}

The sum of the dielectric part (\ref{ediel}) and the scattering
part (\ref{escat}) gives
\begin{eqnarray}
E \gg T\,, \quad  m_e E / m_p &\ll& T \,:  ~\text{or}~ T \ll E \ll
\frac{m_p}{m_e}\,T \,:
\nonumber\\
 {\cal A}_e(v_p)
&\simeq&
   \frac{e_p^2\,\kappa^2_e}{4\pi}\, 
  \left( \frac{\beta_e m_e}{2\pi} \right)^{1/2}
 \, \frac{v_p}{3} \, \left[
\ln \left( \frac{8 T_e m_{pe}^2}{m_e\hbar^2 \kappa_e^2} \right) 
- \gamma -1 \right] \,. 
\label{eresult}
\end{eqnarray}
Figure \ref{fig:015.04} compares this high-energy approximation with the 
exact result. Figure \ref{fig:016.04} shows that the high and low energy
approximations are quite similar. 

\begin{figure}[t]
\vskip-1cm
\includegraphics[scale=0.45]{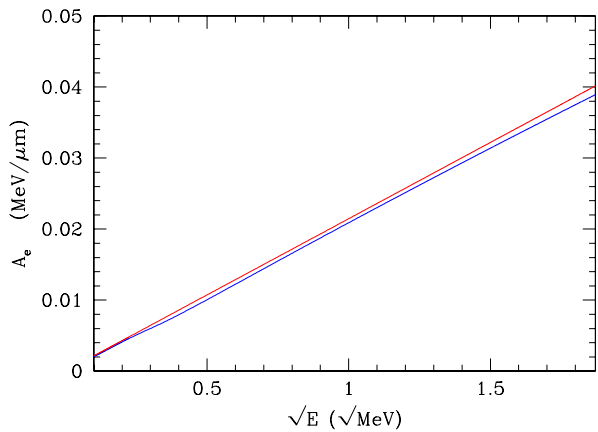}
\vskip-0.8cm 
\caption{\captionskip
 The coefficient $A_e$ (blue curve) compared with the (red curve) 
 high-energy approximation (\ref{eresult}). The plasma is equimolar
 DT with $n_e = 1.0 \times 10^{25} \, {\rm cm}^{-3}$ and 
 $T_e = T_\smI = 10 \, {\rm keV}$. 
 } 
\label{fig:015.04}
\end{figure}
\begin{figure}[b!]
\vskip-1cm
\includegraphics[scale=0.45]{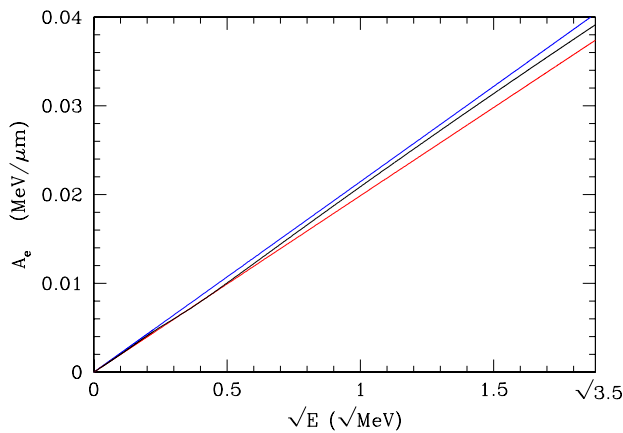}
\vskip-0.8cm 
\caption{\captionskip
  An $\alpha$ particle projectile moving in a equimolar $DT$ plasma 
  with $T_e = T_\smI = 10$ keV and 
  $n_e = 1.0 \times 10^{25} \, {\rm cm}^{-3}$. 
  The (blue curve) low energy approximation (\ref{Lowe}) lies above 
  the (black curve) exact result while the (red curve) high energy  
  approximation (\ref{eresult}) lies below the (black curve) exact 
  result.  Because $\kappa_\smD^2 = 2 \kappa_e^2$ for our equimolar 
  $DT$ plasma, the two approximate forms (\ref{Lowe}) and (\ref{eresult}) 
  differ only by a factor of two inside the logarithm, and this leads 
  to only slightly different slopes.  
  }
\label{fig:016.04}
\end{figure}

Again, to the rough, logarithmic accuracy that produced Eq.~(\ref{Lion})
for the ions, we now have for the electrons
\begin{eqnarray}
E \gg T \,:  \qquad\qquad
{\cal A}_e(v_p) \simeq \frac{e_p^2 \, \kappa_e^2}{4 \pi} \,
        \left(\frac{m_e}{m_p} \, \frac{E}{T_e} \right)^{1/2} \, L \,.
\end{eqnarray}
Note that this electron contribution has the leading temperature
dependence given by the factor 
$ \kappa_e^2 \beta_e^{1/2} \sim T_e^{-3/2} $ 
and thus increases as the temperature is lowered.  This is in marked 
contrast with the corresponding ion contribution given by Eq.~(\ref{IIII}) 
or Eq.~(\ref{V}) which, to within logarithmic accuracy, is independent
of the temperature. The ions dominate at low projectile speeds as shown in
Eq.~(\ref{low}),  and  their contribution at the low speeds also behaves 
as $T_\smI^{-3/2}$ and so also increases as the temperature is lowered.  
On the other hand, as noted immediately below, the electrons greatly 
dominate at very high projectile speeds with a result that is completely 
independent of plasma temperatures. These remarks provide a qualitative 
description of the stopping power behavior in a plasma.

\subsection{$\bm{E \gg m_p T/m_e}$: Electronic Contribution } 

The high velocity limit in this case  has already been calculated by 
BPS in Eq.~(10.43), which we simply quote here:
\begin{eqnarray}
v_p \to \infty \,: &&
\nonumber\\
&& 
{\cal A}_e(v_p) = \frac{e_p^2}{4\pi} \, \frac{\omega_e^2}{v_p^2} \,
     \ln\left(\frac{ 2m_{pe} v_p^2}{\hbar \omega_e} \right) \,.
\label{eeeeee}
\end{eqnarray}
This limit is mostly academic, since the system enters the
relativistic regime at these high velocities. 

\subsection{Energy Cross Over}
\label{crossover}

As we have made explicit, the energy loss to the
ions in the plasma dominates at low projectile energies while the loss
is to the electrons at high projectile energies.  Here we shall 
estimate the crossover point, the projectile energy at which the two
types of loss mechanisms are comparable. We shall find that this
occurs at a projectile energy that is much greater than a typical
plasma temperature $T$, and so we will assume the limit 
$E \gg T$ in estimating the crossover point. 

For the ions, the $E \gg T$ result (\ref{IIII}) reads 
\begin{eqnarray}
  {\cal A}\smI(v_p) &=& 
  \frac{e_p^2}{4\pi}  \, {\sum}_i \, \frac{\omega_i^2}{ v_p^2} 
\,  \left\{
\ln\left(\frac{16\pi}{e_p e_i \kappa_e}\,\frac{m_{pi}v_p^2}{2}\right) 
 - \gamma - \frac{1}{2} \right\} \,.
\label{poor}
\end{eqnarray}
This holds provided that
\begin{equation}
\eta_{pi}^{-2} = \left(\frac{4 \pi \hbar v_p}{e_p e_i}\right)^2
            = \left(\frac{e^2}{e_p e_i}\right)^2 \,
                \frac{2 E}{\alpha^2 \, m_p c^2} 
            \ll 1 \,.
\end{equation}
To put the total ion contribution in a convenient form, we again
define a ``total ion squared plasma frequency'' by
\begin{equation}
{\sum}_i \, \omega_i^2 = \omega_\smI^2 \,,
\end{equation}
replace the ion charge $e_i$ inside the logarithm by a typical
value $e_\smI$, and write $m_{pi} \simeq m_p/2$ to approximate the 
total ion contribution by
\begin{eqnarray}
  {\cal A}_\smI(v_p) &\simeq&
  \frac{e_p^2}{4\pi}  \, \omega_\smI^2 \, \frac{1}{v_p^2} \left[ 
 \ln \left( \frac{16\pi}{e_p e_\smI \kappa_e} \,\frac{E}{2} \right)
 -  \gamma - \frac{1}{2}  \right] \,.
\label{TheI}
\end{eqnarray}
Note that the only temperature dependence in this result is within
the electron Debye wave number inside the logarithm.  Hence the
result only weakly depends upon the plasma temperatures.

A reasonably good approximation for the crossover projectile speed 
$v_p = v_\smC$ should be obtained by equating the ion result (\ref{TheI}) 
to the electronic result (\ref{eresult}) which we repeat here using 
$ \kappa_e^2 = \beta_e m_e \, \omega_e^2 $:
\begin{eqnarray}
 {\cal A}_e(v_p)
&=&
   \frac{e_p^2}{4\pi} \, \frac{\omega_e^2}{3} \,  
  \left( \frac{2}{\pi} \right)^{1/2} \,
  \left( \frac{m_e}{T_\smC} \right)^{3/2} \, v_p \, \left[
\ln \left( \frac{\sqrt{8 T_e m_e}}{\hbar \kappa_e} \right) 
- \frac{1}{2} \Big( \gamma -1 \Big) \right] \,. 
\label{Thee}
\end{eqnarray}
In equating the ion and electron approximations (\ref{TheI}) and
(\ref{Thee}) we use the crossover energy defined by 
\begin{equation}
E_\smC = \frac{1}{2} m_p v_\smC^2
\end{equation}
to obtain
\begin{eqnarray}
E_\smC^{3/2}  \left[ \ln\left(\frac{\sqrt{8T_e m_e}}{\hbar \, \kappa_e}\right)
   -\frac{1}{2} (\gamma + 1) \right] &=& T_e^{3/2}  
\left(\frac{9\pi}{16}\right)^{3/2} \! \left(\frac{m_p}{m_e}\right)^{3/2} 
\frac{\omega_\smI^2}{\omega_e^2} 
   \left[ \ln\left(\frac{8\pi \, E_\smC}{ e_p  e_\smI  \kappa_e}\right)
    - \gamma - \frac{1}{2} \right] \,.
\nonumber\\
&&
\label{ECeq}
\end{eqnarray}
It is important to note that this crossover point only depends upon the
electron temperature $T_e$. The ion temperature $T_\smI$ is of no relevance
here. 

Note that the results that we have obtained provide an approximate form for
the total ${\cal A}$ coefficient as a function of the energy 
$E = m_p v^2_p / 2$,
${\cal A}(E) = {\cal A}_e(E) + {\cal A}_\smI(E)$, namely
\begin{equation}
{\cal A}(E) = \lambda E^{1/2} \, 
\left[ 1 + \left( \frac{E_\smC}{E} \right)^{3/2} \right] \,,
\end{equation}
where
\begin{equation}
\lambda =  \frac{e_p^2}{4\pi} \, \frac{\omega_e^2}{3} \,  
  \left( \frac{1}{\pi} \right)^{1/2} \,
  \left( \frac{m_e}{T_\smC} \right)^{3/2} \, \frac{2}{m_p^{1/2}} \, \left[
\ln \left( \frac{\sqrt{8 T_e m_e}}{\hbar \kappa_e} \right) 
- \frac{1}{2} \Big( \gamma -1 \Big) \right] \,. 
\end{equation}
\begin{figure}[t]
\includegraphics[scale=0.45]{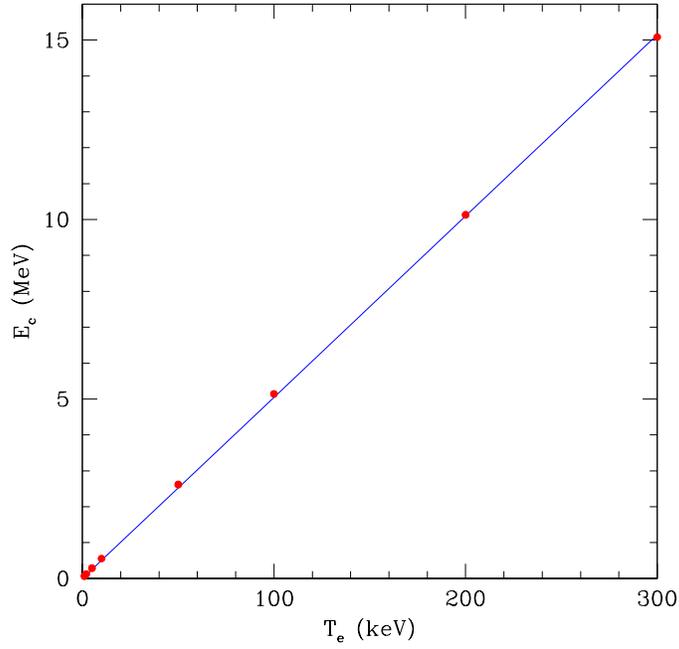}
\vskip 1cm 
\caption{\captionskip
  The solution of the crossover condition (\ref{ECeqN}) as a function
  of the electron temperature for an electron number density
  $n_e = 1.0 \times 10^{25} \, {\rm cm}^{-3}$. The straight line is a
  fit to these points, with $E_\smC = 51 \, T_e$. Similar results obtain     
  for the other densities, with the results presented in Eq.~(\ref{straight}).
}
\label{fig:3}
\end{figure}

To return to assess the validity of our approximation for the cross
over energy, we examine equimolar DT plasmas traversed by alpha
particles of mass $m_p = m_\alpha$, charge $e_p = 2e $, and initial
energy $E_0 = 3.54$ MeV produced by DT fusion.  In numerical terms for
this case with the electron temperature $T_e$ and the crossover energy
$E_\smC$ measured in keV, and the electron number density $n_e$
measured in ${\rm cm}^{-3}$, the crossover relation (\ref{ECeq})
appears as
\begin{eqnarray}
E_\smC^{3/2} \, \left[ \ln\left( 5.796 \times 10^{27} \, 
\frac{T_e^2}{n_e}  \right)
   -1.577 \right] &=& 188.1 \,\, T_e^{3/2}  \,
\left[ \ln\left( 2.66 \times 10^{28} \, \frac{T_e}{n_e} \, E_\smC^2 \right)
    - 2.154  \right] \,.
\nonumber\\
&&
\label{ECeqN}
\end{eqnarray}

\begin{figure}[t]
\includegraphics[scale=0.45]{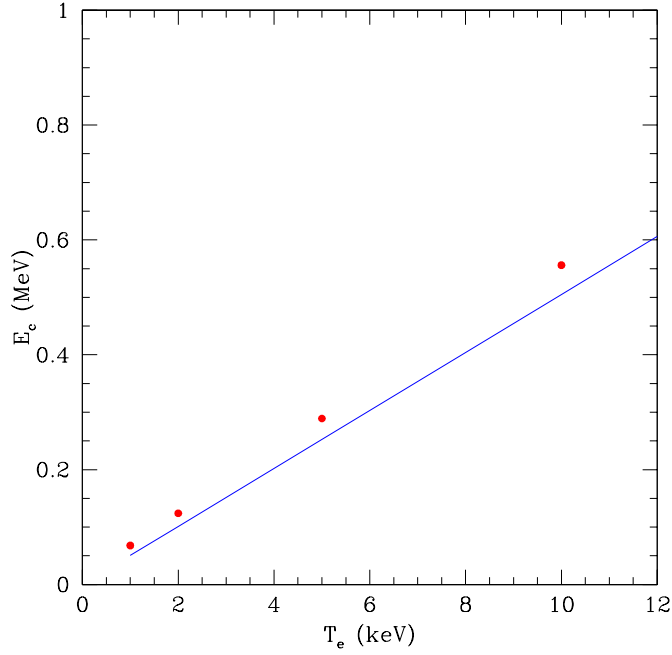}
\vskip 1cm 
\caption{\captionskip
  This figure displays the lower temperature region of the previous 
  Fig.~\ref{fig:3}. Here the linear relation can deviate from the 
  points which are solutions of the crossover
  relation (\ref{ECeqN}) with an error on the order of 10\%.
}
\label{fig:4}
\end{figure}

As shown in Fig.~\ref{fig:3} for one electron number density, we find that
the crossover energies for different electron densities $n_e$ are nearly 
linear functions of $T_e$ given by
\begin{equation} 
E_\smC \simeq T_e \times \left\{
\begin{array}{cc}
48 \,, \quad n_e = 1.0 \times 10^{24} \, {\rm cm}^{-3} \,, \\
51 \,, \quad n_e = 1.0 \times 10^{25} \, {\rm cm}^{-3} \,, \\
53 \,, \quad n_e = 1.0 \times 10^{26} \, {\rm cm}^{-3} \,.
\end{array}
\right.
\label{straight}
\end{equation}
Figure~\ref{fig:4} shows that for energies below 10 keV or so,  these 
linear relations break down by about 10 percent.

\section{The $G$ Function Simplified}
\label{hack}

Here we turn to the definition (\ref{GGdef}) of $G(T_\smI,T_e;E_0)$ in
order to reduce it to a more manageable form. For convenience, we repeat 
this definition here\footnote{We recall that $G$ gives a contribution
$$
\frac{\Delta E_\smI}{E_0} = \left[\frac{ T_e - T_\smI}{E_0} \right] \, 
G(T_\smI,T_e;E_0) \,.
$$
Since the prefactor multiplying $G$ involves a temperature difference
that is at most only 100 keV and the energy $E_0$ is typically 3.5 MeV, 
this prefactor is less than about 3 \%.  Hence to within an accuracy of
a few tenths of \%, we need only compute the pure number $G$ to an
absolute  precision of 0.1 .}:
\begin{eqnarray}
G(T_\smI,T_e;E_0) &=&
\int_0^\infty dE \, F(E) \, e^{-S(E)} 
\nonumber\\
&& \!\!\!\!\!
\int_0^E \frac{dE'}{E'} \,
\frac{e^{ + S(E') }}{ \langle T{\cal A}(E') \rangle } \,
 \left\{  \theta\left( E_0 - E' \right) \, 
 -    \int_{E'}^\infty dE'' \sqrt{E''}  \, 
  \overline{\cal N} \, e^{-S(E'')}  \right\} \,,
\nonumber\\
&&
\label{Fdeff}
\end{eqnarray}
where
\begin{equation}
\overline{{\cal N}}^{\, -1} = \int_0^\infty dE'\,\sqrt{E'}\, e^{-S(E')} \,,
\end{equation}
and
\begin{equation}
F(E) = E \, \frac{{\cal A}_\smI(E) { \cal A}_e(E)} 
              {\langle T {\cal A}(E) \rangle } \,.
\label{F}
\end{equation}

First we note that if $E' < E_0$, the theta function in the curly braces is 
unity. Hence we can make use of the sum rule (\ref{summ}) to write  
\begin{equation}
 G(T_\smI,T_e;E_0) =  G_1(T_\smI,T_e;E_0)+ G_2(T_\smI,T_e;E_0)+
 G_3(T_\smI,T_e;E_0) \,, 
\label{gparts}
\end{equation}
where
\begin{eqnarray}
&& G_1(T_\smI,T_e;E_0) = 
\int_0^{E_0} dE \, F(E) \,  e^{-S(E)} 
\, \int_0^E \frac{dE'}{E'} \,
\frac{e^{ + S(E') }}{ \langle T{\cal A}(E') \rangle } \,
 \int_0^{E'} dE'' \sqrt{E''} \, \,
  \overline{\cal N} \, e^{-S(E'')} \,,
\nonumber\\
&&
\label{gone}
\end{eqnarray}
\begin{eqnarray}
&& G_2(T_\smI,T_e;E_0) = 
\int_{E_0}^\infty dE \, F(E) \, e^{-S(E)} 
\, \int_0^{E_0} \frac{dE'}{E'} \,
\frac{e^{ + S(E') }}{ \langle T{\cal A}(E') \rangle } \,
 \int_0^{E'} dE'' \sqrt{E''} \, \,
  \overline{\cal N} \, e^{-S(E'')}   \,,
\nonumber\\
&&
\label{gtwo}
\end{eqnarray}
and
\begin{eqnarray}
&& G_3(T_\smI,T_e;E_0) = 
 -
\int_{E_0}^\infty dE \, F(E) \,  e^{-S(E)} 
\int_{E_0}^E \frac{dE'}{E'} \,
\frac{e^{ + S(E') }}{ \langle T{\cal A}(E') \rangle } \,
  \int_{E'}^\infty dE'' \sqrt{E''} \, \,
  \overline{\cal N} \, e^{-S(E'')} \,.
\nonumber\\
&&
\label{gthree}
\end{eqnarray}

First we show that $G_3$ may be neglected.
For the very last pair of integrals in $G_3$, since the energies $E'$ and
$E''$ are larger than $E_0 \gg T_e \,,\, T_\smI$, the electron 
contribution to the ${\cal A}_b$ functions dominate, and so
\begin{equation}
S(E'') - S(E') = \frac{1}{T_e} \, \left( E'' - E' \right) \,.
\label{SBig}
\end{equation}
This is a very large number unless $E''$ is near $E'$.  Hence,
with corrections that will be of the very small order 
$T_e / E_0 $, we have
\begin{eqnarray}
&& \int_{E_0}^E \frac{dE'}{E'} \,
\frac{e^{ + S(E') }}{ \langle T{\cal A}(E') \rangle } \,
  \int_{E'}^\infty dE'' \sqrt{E''} \, \, 
  \overline{\cal N} \, e^{-S(E'')}  
\nonumber\\
&& \qquad\qquad \simeq 
\int_{E_0}^E \frac{dE'}{E'} \,
\frac{1}{ T_e {\cal A}_e(E')  } \,
  \sqrt{E'} \, \,   \overline{\cal N} \, 
\int_{E'}^\infty dE'' 
\exp\left\{ - \frac{1}{T_e} \, \left( E'' - E' \right) \right\} 
\nonumber\\
&& \qquad\qquad =
   \overline{\cal N} \, 
\int_{E_0}^E \frac{dE'}{\sqrt E'} \,
\frac{1}{{\cal A}_e(E')  } \,,
\end{eqnarray}
and so, again since the electrons dominate the ${\cal A}_b$ functions in
the high-energy regions that appear here, 
\begin{eqnarray}
&&  G_3(T_\smI,T_e;E_0) 
 \simeq - 
\int_{E_0}^\infty dE \, \frac{E}{T_e} \,
  {\cal A}_\smI(E) \,  e^{-S(E)} \,
     \overline{\cal N} \, 
\int_{E_0}^E \frac{dE'}{\sqrt E'} \,
\frac{1}{{\cal A}_e(E')  } \,.
\end{eqnarray}
There is really no need to go any further in the evaluation of 
$ G_3(T_\smI,T_e;E_0) $ since it has the exponentially small factor
$\exp\{-S(E)\}$, with $E\geq  E_0$. In this region, as we have noted,
the electrons dominate and so 
$\exp\{ -S(E) \} \simeq \exp\{ - E / T_e \} $.  
Even for an electron temperature as high as 35 keV and for a $DT$ 
fusion alpha particle with $E_0 = 3.54$ MeV,  this factor is 
$\exp\{ - 100 \} \simeq 4 \times 10^{-44} $.

For the evaluation of $G_2$, it is convenient to define
\begin{equation}
H(E') = 
\int_0^{E'} dE''  \sqrt{E''}  \, 
  \overline{\cal N} \, e^{-S(E'')}   \,.
\label{H}
\end{equation}
To isolate the leading pieces, we shall write
\begin{equation}
e^{\pm S(E)}  =
\pm \frac{\langle T {\cal A}(E) \rangle}{{\cal A}(E)} \, 
\frac{d}{d E} \, e^{\pm S(E)} 
\label{pm}
\end{equation}
and integrate by parts.  This will provide an extra explicit factor 
of a plasma temperature $T$ in the numerator, thereby yielding a 
small quantity.  

The final double integral in the triple integral (\ref{gtwo}) defining 
$G_2$ now appears as
\begin{eqnarray}
&& \int_0^{E_0} \frac{dE'}{E'} \,
\frac{e^{ + S(E') }}{ \langle T{\cal A}(E') \rangle } \,
 \int_0^{E'} dE'' \sqrt{E''} \, 
  \overline{\cal N} \, e^{-S(E'')}   
\nonumber\\
&& \qquad\qquad =
\int_0^{E_0} dE' \,  \, \frac{H(E')}{E' \, {\cal A}(E')} 
\frac{d}{d E'} \, e^{ + S(E') }
\nonumber\\
&& \qquad\qquad \simeq
 \frac{H(E_0)}{E_0 \, {\cal A}(E_0) } \, e^{ + S(E_0) }
- \int_0^{E_0} dE' \, e^{ + S(E') }
\frac{d}{d E'} \, \left[  \frac{H(E')}{E' \, {\cal A}(E') } \right]
\nonumber\\
&& \qquad\qquad \simeq
 \frac{H(E_0)}{E_0 \,  {\cal A}(E_0) } \, e^{ + S(E_0) }
- \frac{\langle T {\cal A}(E_0) \rangle}{{\cal A}(E_0)} \, 
e^{ + S(E_0) }
\left. 
\frac{d}{d E} \, \left[ \frac{H(E)}{E \, {\cal A}(E) } \right] 
\right|_{E_0}  + \cdots \,,
\nonumber\\
&&
\label{series}
\end{eqnarray}
where the ellipsis represents the series resulting by further partial 
integrations.  As we shall see, the second term in the last line of
Eq.~(\ref{series}) is already negligible, and so are these omitted terms.
The approximate equalities in Eq.~(\ref{series}) neglect lower limit terms 
since they result in exponentially small quantities from the remaining 
integration over $E$ in Eq.~(\ref{gtwo}) because of the factor 
$\exp\{-S(E)\}$ with $E > E_0$. Here, to within very good accuracy,
\begin{equation}
H(E_0) = 1 \,,
\end{equation}
since the integral defining $H(E)$ has long since converged to 
its limiting value at $E=E_0$. Hence,
\begin{eqnarray}
&& \int_0^{E_0} \frac{dE'}{E'} \,
\frac{e^{ + S(E') }}{ \langle T{\cal A}(E') \rangle } \,
 \int_0^{E'} dE'' \sqrt{E''} \, 
  \overline{\cal N} \, e^{-S(E'')}   
\nonumber\\
&& \qquad\qquad \simeq
 \frac{1}{E_0 \,  {\cal A}(E_0) } \, e^{ + S(E_0) } \left\{ 1
- \langle T {\cal A}(E) \rangle \left. 
E \, \frac{d}{d E} \, \left[ \frac{1}{E \, {\cal A}(E) } \right] 
\right|_{E_0} \right\} \,.
\label{gtwoo}
\end{eqnarray}
Here, since at large energies the rate of energy variation is of
order $1/E$,
\begin{equation}
 \langle T {\cal A}(E) \rangle \left. 
E \, \frac{d}{d E} \, \left[ \frac{1}{E \, {\cal A}(E) } \right]
\right|_{E_0} 
\sim \, \frac{ \langle T {\cal A}(E_0) \rangle }{ E_0 \, {\cal A}(E_0) } 
\, \sim \, \frac{T}{E_0} \,,
\label{more}
\end{equation}
in which $T$ is a typical plasma temperature. The ratio $T / E_0 $ is at
most a few percent for the plasma configurations that we consider, and
thus it is a good approximation to replace the curly braces in 
Eq.(\ref{gtwoo}) by unity. 

Recalling the definition (\ref{F}) of $F(E)$ and then using the relation 
(\ref{pm}), we obtain
\begin{eqnarray}
G_2(T_\smI,T_e;E_0) &\simeq& 
\frac{1}{E_0 \, {\cal A}(E_0)  } \, 
\int_{E_0}^\infty dE \, 
\left\{ E \, \frac{{\cal A}_\smI(E) { \cal A}_e(E)} 
              {\langle T {\cal A}(E) \rangle } \right\} \,
\exp\left\{-\left[S(E) - S(E_0)\right]\right\} 
\nonumber\\
&=& 
 - \frac{1}{E_0 \, {\cal A}(E_0)}  \, 
\int_{E_0}^\infty dE \, E \,  
\frac{{\cal A}_\smI(E) {\cal A}_e(E)}{{\cal A}(E)} \, 
\frac{d}{dE} \,  \exp\left\{-\left[S(E) - S(E_0)\right]\right\} 
\nonumber\\
&=& 
 \frac{{\cal A}_\smI(E_0) \, {\cal A}_e(E_0)}{{\cal A}^2(E_0) } 
\nonumber\\
&&  +
\frac{1}{E_0 \, {\cal A}(E_0)}  \, 
\int_{E_0}^\infty dE \,\exp\left\{-\left[S(E) - S(E_0)\right]\right\} 
\, \frac{d}{dE} \, \left\{ E \,  
\frac{{\cal A}_\smI(E) {\cal A}_e(E)}{{\cal A}(E)} \right\} \,. 
\nonumber\\
&&
\label{geetwo}
\end{eqnarray}
As before, we have the estimate
\begin{eqnarray}
 \frac{d}{dE} \, \left\{ E \,  
\frac{{\cal A}_\smI(E) {\cal A}_e(E)}{{\cal A}(E)} \right\} 
&\sim&    
\frac{{\cal A}_\smI(E) {\cal A}_e(E)}{{\cal A}(E)}  
= \frac{\langle T {\cal A}(E) \rangle}{ E \,{\cal A}(E)} \,
  \left\{ E \,  
\frac{{\cal A}_\smI(E) {\cal A}_e(E)}{\langle T {\cal A}(E)\rangle} \right\} 
\nonumber\\
&\sim& \frac{T}{E} \,   \left\{ E \,  
\frac{{\cal A}_\smI(E) {\cal A}_e(E)}{\langle T {\cal A}(E)\rangle} \right\} \,.
\label{est}
\end{eqnarray}
Here again $T$ represents a typical plasma temperature, and since the
integration region starts at $E = E_0$, we have $T / E \leq T / E_0 $,
Since the factor in the curly braces in the last line in Eq.~(\ref{est}) 
is just the factor in the curly braces in the first line in Eq.~(\ref{geetwo}),
we see that the last line in Eq.~(\ref{geetwo}) is of order $ T / E_0 $
times the first line, and thus gives a correction on the order of a few 
percent. We have found that, to within corrections of a few percent,
\begin{eqnarray}
G_2(T_\smI,T_e;E_0) &\simeq& 
 \frac{{\cal A}_\smI(E_0) \, {\cal A}_e(E_0)}{{\cal A}^2(E_0) } \,.
\label{geee2}
\end{eqnarray}
The accuracy of the analytical approximation (\ref{geee2}) for $G_2$
has been confirmed to this precision by direct numerical evaluation 
of its definition (\ref{gtwo}).

In summary, Eq.~(\ref{gparts}) expresses the $G$ function in three
parts.  The first part $G_1$ involves a triple integral that must be
evaluated by numerical computation. This evaluation is simplified
because, with the partition that we have made, the regions of
integration that appear in $G_1$ are restricted to the finite interval $0
< E < E_0$. For the second part $G_2$, the approximation (\ref{geee2})
is sufficiently accurate for our purposes.  The remainder $G_3$ is
very small and we may simply set
\begin{equation}
G_3(T_\smI,T_e;E_0) = 0 \,.  
\end{equation} 

\pagebreak

\end{document}